\documentclass[epsfig,11pt]{article}

\usepackage{epsfig}
\usepackage{graphicx}
\usepackage{array}
\usepackage{caption}
\usepackage{subcaption}
\usepackage{slashed}
\usepackage{amsmath}
\usepackage{amssymb}
\usepackage{braket}
\usepackage{comment}

\newcommand{\beq}{\begin{equation}}   
\newcommand{\eeq}{\end{equation}}
\newcommand{\beqn}{\begin{eqnarray}}   
\newcommand{\eeqn}{\end{eqnarray}}

\newcommand{\bea}{\begin{eqnarray}}
\newcommand{\eea}{\end{eqnarray}}
\newcommand{\be}{\begin{equation}}
\newcommand{\ee}{\end{equation}}
\newcommand{\bead}{\begin{aligned}}
\newcommand{\eead}{\end{aligned}}

\newcommand{\nn}{\nonumber}

\newcommand{\gsim}{\lower.7ex\hbox{$
\;\stackrel{\textstyle>}{\sim}\;$}}
\newcommand{\lsim}{\lower.7ex\hbox{$
\;\stackrel{\textstyle<}{\sim}\;$}}
\begin{document}

\begin{titlepage}

\begin{flushright}
FTPI-MINN-15/39

UMN-TH-3449/15
\end{flushright}

\vspace{0.7cm}

\begin{center}
{  \large \bf  Spin vortices in the Abelian-Higgs model with cholesteric vacuum structure}
\end{center}
\vspace{0.6cm}

\begin{center}
 {\large 
Adam Peterson,$^a$ Mikhail Shifman,$^{a,b}$ and Gianni Tallarita$^{c,d}$}
\end {center}

\vspace{3mm}
 
\begin{center}

$^a${\em University of Minnesota, School of Physics and Astronomy,
Minneapolis, MN 55455, USA}\\[1mm]
$^b${\em William I. Fine Theoretical Physics Institute, University of Minnesota,
Minneapolis, MN 55455, USA}\\[1mm]
$^c${\em Departamento de Ciencias, Facultad de Artes Liberales, Universidad Adolfo Ib\'{a}\~{n}ez, Santiago 7941169, Chile}\\[1mm]
$^d${\em Centro de Estudios Cient\'{i}ficos (CECs), Casilla 1469, Valdivia, Chile}
\\[1mm]

\end {center}

\vspace{2cm}

\begin{center}
{\large\bf Abstract}
\end{center}

We continue the study of $U(1)$ vortices with cholesteric vacuum structure. A new class of solutions is found which represent global vortices of the internal spin field. These spin vortices are characterized by a non-vanishing angular dependence at spatial infinity, or winding.  We show that despite the topological $\mathbb{Z}_2$ behavior of $SO(3)$ windings, the topological charge of the spin vortices is of the $\mathbb{Z}$ type in the cholesteric.  We find these solutions numerically and discuss the properties derived from their low energy effective field theory in $1+1$ dimensions.

\hspace{0.3cm}

\end{titlepage}


\section{Introduction}
\label{intro}

This paper is a continuation of \cite{Peterson:2014nma} which investigates the low energy theory of $U(1)$ vortices in systems with cholesteric vacuum structures.  This work was inspired by considerations from supersymmetric solitons (see \cite{Shifman:2009} for a review) as well as superconductivity \cite{Carlson:2003} and superfluidity \cite{Salomaa:1987}.  The kind of vortex solutions found in \cite{Peterson:2014nma} are similar to the $w$ vortices appearing in superfluid $^3$He-B \cite{Salomaa:1987}.  Indeed these $w$ vortices, and those found in \cite{Peterson:2014nma}, develop a cholesteric spiral structure in the core.  If a collection of such vortices are coherently orientated in a rotating cryostat the spiral structure appears in the bulk superfluid \cite{Salomaa:1987}. The main task of this paper is to report on a different class of solutions also present in this model. These solutions are characterized by the internal spin field carrying a topological charge of its own.  Typically vortices with topological windings originating from an internal $SO(3)_J$ spin group are referred to as spin vortices.  However, for reasons discussed below, the solutions we obtain are distinct from the typical spin vortex observed in say superfluid $^3$He \cite{Volovik:1977}\cite{Kondo:1992}.  Our plan is to find these solutions numerically and investigate their low energy excitations.

The model, as originally introduced in \cite{Peterson:2014nma}, has a symmetry structure given by
\begin{equation}
G = U(1)_{\rm gauge} \times SO(3)_{J} \times T
\end{equation}
where the $U(1)_{\rm gauge}$ group represents gauge rotations of a complex Higgs field, and $SO(3)_{S+L} \equiv SO(3)_J$ represents the spin orbit locked rotational symmetry.  The group $T$ represents the three translational symmetries of the model.  The spin orbit locked $SO(3)_{J}$ appears from the larger group $SO(3)_S \times SO(3)_L$ due to an additional term in the Largrangian coupling the spin index to the spatial gradiant.  This term is parity violating and leads to spontaneous breaking of translational symmetry.  The additional term thus gives rise to a cholesteric spiral structure in the vacuum under certain conditions of the parameters in the model \cite{Peterson:2014nma}. 

The spontaneous breaking of the $U(1)_{\rm gauge}$ group allows for the appearence of the Abrikosov-Nielsen-Olsen \cite{Abrikosov:1957} flux tube familiar to models of type II superconductors \cite{Carlson:2003} and Yang-Mills theories \cite{Shifman:2009}.  With no other winding of the subgroups of $G$ these vortices represent the mass vortices that were studied in \cite{Peterson:2014nma}.  A typical spin vortex on the other hand arises from the non-trival topology of the $SO(3)_{J}$ group:
\begin{equation}
\pi_1(SO(3)_{J}) = \mathbb{Z}_2.
\end{equation}
The spin vortices can appear regardless of the cholesteric behavior of the vacuum and only require the existence of a broken $SO(3)$ symmetry.  However, the vortices we consider in this paper are distinct from the typical spin vortex in that they are only allowed due to the broken translational symmetry of the vacuum.  In particular, the vacuum of our model preserves a $U(1)_{J'_z}$ global subgroup of $G$ which is generated by a linear combination of the translational and rotational generators about an axis oriented along the cholesteric wave vector $\vec{k}$:
\begin{equation}
G \rightarrow U(1)_{J'_z}\times T_{x,y}, \mbox{ where } U(1)_{J'_z} \subset U(1)_{J_z} \times T_z.
\end{equation}
$T_{x,y}$ is the translational group along the direction perpendicular to $\vec{k}$, whilst $T_z$ is that along $\vec{k}$.  Here the subgroup $U(1)_{J'_z}$ is denoted with its generator $J'_z$ which is a linear combination of the rotational $J_z$ and translational $p_z$ generators:
\begin{equation}
J'_z \equiv J_z-\frac{p_z}{k}, \;\;\; k \equiv |\vec{k}|.
\end{equation}

As we will discuss below this equivalence of rotations and translations in the vacuum leads to a global $U(1)$ degeneracy space of vacua in addition to the $U(1)_{\rm gauge}$ ``degeneracy" appearing from the Higgs phase.  Vortices may appear with either gauge or global $U(1)$ topological windings or both:
\begin{equation}
\pi_1(U(1)_{\rm gauge} \times U(1)_{}) = \mathbb{Z} + \mathbb{Z}.
\end{equation}

In this paper we will focus our attention on the vortices with both types of charges.  We hasten to point out that the vortices in this model are global.  Thus, for cholesteric vacua (vacuum $I$ as it is defined below) these vortices will have a logarithmically divergent energy functional and thus require an infrared cutoff.  This is not a problem for systems in a finite volume such as superfluid $^3$He.  However, as a consequence of this finite volume, the Kelvin modes appearing on the vortex will acquire a mass gap proportional to the inverse cutoff length $R$.  These effects were considered recently in \cite{Kobayashi:2013gba}, and we will apply similar arguments when we discuss the gapless excitations in section 5 below.

We will begin in sections 2 and 3 with a detailed review of the model and the vacuum structure following the procedure discussed in \cite{Peterson:2014nma}.  We will explore briefly the Goldstones appearing in the bulk cholesteric vacuum.  We will discuss their relation to the spin vortices appearing in the cholesteric bulk.  In section 4 we will present the equations of motion for the spin vortices and present their numerical solutions for a representative set of parameters.  In section 5 we will discuss the low energy theory of the moduli appearing on the vortices.  We will present an analysis of the Kelvin modes for vortices in a particular vacuum, and provide additional topological discussions of the orientational moduli.  For the case of a cholesteric vacuum we will discuss the finite volume effects on the mass terms for Kelvin modes.  Finally, we provide concluding remarks in section 6.

\section{The system}

The starting point is the system used in \cite{Peterson:2014nma}. This model is an extension of the model originally suggested in \cite{Shifman:2013}, inspired by Witten's superconducting string model \cite{Witten:1985}, and further studied in \cite{Shifman:2013a}. Its Lagrangian has the form
\begin{align} 
\mathcal{L}  &=\mathcal{L}_0+\mathcal{L}_\chi+\mathcal{L}_\varepsilon, \notag \\[2mm]
\mathcal{L}_0 &= -\frac{1}{4e^2}F_{\mu\nu}F^{\mu\nu}+|D_\mu \phi|^2-\lambda(|\phi|^2-v^2)^2, \notag \\[2mm]
\mathcal{L}_\chi &=\partial_t\chi_i\partial_t \chi_i - \partial_i \chi_j \partial_i \chi_j-\gamma \left[(-\mu^2+|\phi|^2)\chi_i\chi_i+\beta(\chi_i\chi_i)^2\rule{0mm}{4mm}\right], \notag \\[2mm]
\mathcal{L}_\varepsilon &= -\eta \varepsilon_{ijk}\chi_i \partial_j \chi_k\,,
\label{Lagrangian}
\end{align}
where the subscript $i$ runs over $i=1,2,3$. The field $\chi_i$ can be viewed as a spin field. The covariant derivative is defined in a standard way
\begin{equation}
D_\mu = \partial_\mu- i A_\mu\,.
\end{equation}
The last term in (\ref{Lagrangian}) is the parity violating twist term with a single derivative.  This term is responsible for the cholesteric behavior of the vacuum as we will show below.

Let us consider the system with $\eta=0$ first. Then, under an appropriate choice of parameters the charged field $\phi$  condenses in the ground state,
\begin{equation}
|\phi|_{\rm vac} = v\,,
\end{equation}
and, if $\mu<v$, the field $\chi$ is not excited in the vacuum
\begin{equation}
\left(\chi_i\right)_{\rm vac} = 0\,.
\end{equation}
The system has full translational invariance $T$ and orbital rotational symmetry $SO(3)_L$ as well as an internal rotational $SO(3)_S$ symmetry of the $\chi_i$ spin field. As is well known, the model supports the Abrikosov flux tube \cite{Abrikosov:1957}. Inside the tube, where the Higgs field vanishes, the spin field $\chi_i$ is excited giving rise to gapless (or quasigapless) excitations of non-Abelian type, localized on the vortex. The full solution is known as a non-Abelian string.\newline
 
Including $\mathcal{L}_\varepsilon$, explicitly breaks the orbital rotational part of the Lorentz symmetry  implying a spin-orbit locked symmetry of the full Lagrangian,
\begin{equation}
SO(3)_L \times SO(3)_S \rightarrow SO(3)_{S+L} \equiv SO(3)_J. 
\end{equation}
The energy density derived from the Lagrangian (\ref{Lagrangian}) is
\begin{align}
E = &\frac{1}{4e^2}F_{ij}F^{ij}+|D_i\phi|^2+\lambda(|\phi|^2-v^2)^2+\partial_i\chi_j\partial_i\chi_j \nonumber\\[2mm]
&+ \eta \varepsilon_{ijk}\chi_i \partial_j \chi_k+\gamma \left[(-\mu^2+|\phi|^2)\chi_i\chi_i+\beta(\chi_i\chi_i)^2\right].
\label{EnergyDensityEta}
\end{align}
Note that the kinetic terms of $\chi_i$ including the linear in derivatives $\eta$ term can be recognized as the Frank-Oseen free energy density of an isotropic chiral nematic liquid crystal \cite{deGennes:1993}. 

For later convenience we rescale all couplings in the model to make them dimensionless. Hence, we take
\begin{equation}
\tilde{z} = m_\phi z\,,
\end{equation}
to denote the distance in the longitudinal direction aligned with the string axis.  We will also set the dimensionless radial distance
\begin{equation}
\rho= m_\phi\sqrt{x^2+y^2}\,,
\end{equation}
where the mass of the elementary excitations of the charged field $\phi$  is \be m_{\phi}^2 = 4\lambda v^2. \ee  Additional dimensionless parameters are given as
\begin{equation}
b= \frac{\gamma(c-1)}{4\lambda c}\,,\quad c=\frac{v^2}{\mu^2}\,,\quad  a=\frac{e^2}{2\lambda}\,, \quad \tilde{\eta}=\eta/m_\phi\,.
\end{equation}
We limit ourselves to the semi-classical approximation and assume all couplings in the model are small. \newline

The field $\chi_i$ being represented in Cartesian coordinates takes the form
\begin{equation}
\chi_i = \frac{\mu}{\sqrt{2\beta}}\, \left\{ \rule{0mm}{4mm} \tilde{\chi}_x(x,y,z),\,\, \tilde{\chi}_y(x,y,z),\,\, \tilde{\chi}_z(x,y,z) \right\}.
\end{equation}

The static classical equations of motion are derived by extremization of energy (\ref{EnergyDensityEta}), in a general coordinate system they read
\begin{align}
\label{eomm}
& \nabla_i\left(\sqrt{-g}g^{ij}\nabla_j \chi_k\right)-\eta\varepsilon_{kji}\nabla_j\chi_i-\frac{\partial V}{\partial \chi_k}=0\,,\nonumber\\[3mm]
& D_i\left(g^{ij}\sqrt{-g}D_j\phi\right)+\sqrt{-g}\left(2\lambda\left(|\phi|^2-v^2\right)+\gamma g^{ij} \chi_i\chi_j\right)\phi=0\,,\nonumber\\[4mm]
& \partial_i\left(\sqrt{-g}g^{ji}g^{kl}F_{jk}\right)-ie^2\sqrt{-g}\left(\phi^*D^l\phi-\phi D^l\phi^*\right)=0\,,
\end{align}
where
\begin{equation}
\frac{\partial V}{\partial \chi_k} = \sqrt{-g}\gamma\left( \left(-\mu^2+|\phi|^2\right)+2\beta g^{ij}\chi_i\chi_j\right)\chi_k\,,
\end{equation}
and 
\begin{equation}
\nabla_i\chi_j = \partial_i\chi_j - \Gamma^k_{ij}\chi_k
\end{equation}
is the standard curved space covariant derivative with $\Gamma$ the Christoffel symbol.

\section{Ground state}

In this section we give a summary of the results obtained in \cite{Peterson:2014nma} concerning the ground state structure of the model.  As shown there this model has a rich ground state structure depending on the value of $\eta$ and the relation between the parameters. Given the presence of the first derivative term one expects that in the ground state translational invariance will be spontaneously broken. If one assumes that this breaking is aligned in the $z$ direction, then a cholesteric structure appears in the ground state with the $\chi$ field rotating in the $(x,y)$ plane
\bea
\chi_i &=& \chi_0\, \epsilon_i(z)\,,\qquad \frac{\sqrt{2\beta}}{\mu}\, \chi_i = \tilde\chi_0\, \epsilon_i(z)\,,
\nonumber\\[2mm]
{\vec\epsilon}\,(z) &=& \left\{ \rule{0mm}{4mm} \cos{kz},\,\,\sin{kz},\,\, 0\right\}\,,
\label{GroundState}
\eea
where $k$ is the $z$ component of the wave vector $\vec k$.  $\chi_0$ is the constant amplitude of the $\vec{\chi}$ field in the vacuum to be determined below in (\ref{20}).  Then, in the ground state we may take
\be
\phi = \phi_0\,, \,\,\, \quad A_i=0\,,
\ee
where $\phi_0 $ is a constant.  The equations of motion (\ref{eomm}) imply
\bea
&& \chi_0\left(-k^2+\eta k-\gamma((-\mu^2+ \phi_0^2)+2\beta\chi_0^2)\right)=0\,,
\nonumber\\[2mm]
&&\phi_0\left(2\lambda(\phi_0^2-v^2)+\gamma\chi_0^2\right)=0\,.
\label{20}
\eea
Minimizing over the wavevector $k$ one obtains
\be
k=\eta/2. \label{keta}
\ee
Therefore, only when $\eta=0$ we recover translational invariance (however see (\ref{Vac2}) below).
 As per \cite{Peterson:2014nma} we always assume
\be
c >1 ,\quad b >0, \quad \beta > 0\quad \text{and} \quad a>0,
\ee
and we use here the same dimensionless field definitions
\bea
\tilde{E}_{vac} &=& \frac{m_\phi}{v^2}E_{vac}\nn\\
\tilde{\chi}_0 &=& \frac{\mu}{\sqrt{2\beta}}\chi_0\nn\\
\varphi &=& v\phi.  
\eea

There are two parameter ranges which characterize the vacuum structure.  We will report here the main results and refer the reader to  \cite{Peterson:2014nma}  for calculational details. For both cases the lowest energy vacuum solution is dictated by the value of $\tilde{\eta}$.  This is important in order to understand the role of the spin-vortex solutions and their winding at spatial infinity. 
Consider first the case where 
\be
\beta(c-1) > bc.
\ee
Then when
\begin{equation}
\quad  4b < \tilde{\eta}^2 < 4\beta\left(1 - \frac{b}{\beta(c-1)}\right),
\label{Condition1}
\end{equation}
the vacuum solution of (\ref{20}) is
\begin{equation}
(\varphi^2_0)^{I} = \frac{1-\frac{b}{\beta (c-1)}-\frac{\tilde{\eta}^2}{4 \beta}}{1-\frac{bc}{\beta (c-1)}}\,,\quad (\tilde\chi^2_0)^{I} = (c-1)\frac{\frac{\tilde\eta^2}{4b}-1}{1-\frac{bc}{\beta (c-1)}} \nonumber,
\label{Vac1}
\end{equation}
\begin{equation}
\tilde{E}_{vac}^{I} =-\frac{(c-1)^2(\tilde{\eta}^2-4 b)^2}{\beta (c-1) -bc}.
\label{E1}
\end{equation}
In this vacuum both $\varphi_0\neq0$ and $\chi_0\neq0$ and therefore both local $U(1)$ gauge and $z-$translational symmetries are broken.  This vacuum allows for the existence of local ANO vortices.
If instead
 \be \tilde{\eta}^2 < \tilde{\eta}_{{\rm crit}_1}^2 \equiv 4b\ee
 then the vacuum solution is
 \begin{equation}
\label{Vac2}
(\varphi^2_0)^{II} = 1,\quad \left(\tilde{\chi}_0\right)^{II}=0, \quad \tilde{E}_{vac}^{II} = 0,
\end{equation}
Note that, since $\left(\tilde{\chi}_0\right)^{II}=0$, this vacuum preserves translational invariance.
Finally, when 
\begin{equation}
\tilde{\eta}^2_{{\rm crit}_2} \equiv 4\beta \left(1-\frac{b}{\beta(c-1)}\right),
\end{equation}
the vacuum solution is given by
\begin{equation}
(\varphi_0)^{III} =0, \quad (\tilde{\chi}^{2}_0)^{III}=1+(c-1)\frac{\tilde{\eta}^2}{4b}, \nonumber\\
\label{Vac3}
\end{equation}
\begin{equation}
\tilde{E}_{vac}^{III} = \frac{16 bc(c-1)\beta-((c-1)\tilde{\eta}^2+8b)^2}{64bc(c-1)\beta}.
\label{E3}
\end{equation}
In this vacuum translational invariance is broken but local $U(1)$ gauge invariance is preserved, hence this vacuum does not support ANO vortices.  The energy of the ground states as functions of $\eta$ are shown in Figure 1.

In the opposite case where
 \be\beta(c-1) < bc,\ee
 the minimizing vacuum is given by (\ref{Vac2}) when
\begin{equation}
\tilde{\eta}^2 < \tilde{\eta}^2_{{\rm crit}_3}=\frac{\sqrt{(c-1)bc\beta}-b}{(c-1)}.
\end{equation}
When $\tilde{\eta}$ crosses the critical value $\tilde{\eta}^2_{{\rm crit}_3}$, there is a first order transition from vacuum (\ref{Vac2}) to vacuum (\ref{E3}).  This pattern of ground states is depicted in Figure 2.

For this paper we will focus our attention primarily on the vacuum $I$ and vacuum $II$ cases where the gauge $U(1)_{\rm gauge}$ group is broken by a non-zero $\phi$ field.  For vacuum $II$ none of the global symmetries $SO(3)_J$ or $T$ are broken in the ground state.  For ANO vortices in this case, cholesteric behavior of the $\chi_i$ field will only appear in the vortex core.  On the other hand vacuum $I$ has an interesting symmetry structure appearing from a linear combination of the rotational $J_z$ and translational $p_z$ generators:
\begin{equation}
J'_z = J_z - \frac{p_z}{k}.
\end{equation}
From (\ref{GroundState}) for non-zero $\chi_0$ it is clear that $J'_z$ annihilates the vacuum, and thus the vacuum $I$ ground state retains a $U(1)_{J'_z}$ symmetry in addition to the subgroup $T_{xy}$ of translations perpendicular to $\vec{k} = k\hat{z}$.  We may thus write the symmetry group of the vacuum $I$
\begin{equation}
H_I = U(1)_{J'_z} \times T_{xy}.
\end{equation}

The equivalence of translations and rotations about the $z$ axis on the vacuum state implies an additional structure on the degeneracy space of vacua in vacuum $I$.  Consider the coset space
\begin{equation}
G/H_I = U(1)_{\rm gauge} \times S^2_{J_\perp} \times U(1)_{J''_z}/\mathbb{Z}_2,
\end{equation}
where
\begin{equation}
J''_z \equiv J_z +\frac{p_z}{k}
\end{equation}
is a generator of the $U(1)_{J''_z}$ group orthogonal to $J'_z$.  Here $S^2_{J_\perp}$ is the two sphere of directions for the wave vector $\hat{k}$.  The group $U(1)_{J''_z}$ generates non-trivial rotations of the vacuum about the wave vector $\vec{k}$.  Physically the group $U(1)_{J''_z}$ represents degeneracy associated with rotations of $\vec{\chi}$ about $\vec{k}$.  Alternatively, due to the equivalence of translations and rotations, and the periodicity of the $z$ translations of the vacuum, the $U(1)_{J''_z}$ may be viewed as a compactification of $T_z$.

Naively the topology of $S^2 \times U(1)/ \mathbb{Z}_2$ supports only a $\mathbb{Z}_2$ winding.  However, it was shown in \cite{Radzihovsky:2011} that for low energy excitations the generators $J_x$ and $J_y$ acquire a mass gap through an effective Higgs mechanism.  For the vacuum state this implies that only a single Goldstone mode associated with phase rotations of the vacuum appear, instead of the naively expected three Goldstones from the three broken generators.  Effectively we may write
\begin{equation}
G/H_I \rightarrow U(1)_{\rm gauge} \times U(1)_{J''_z}.
\end{equation}
The effective $U(1)_{J''_z}$ degeneracy of vacuum $I$ implies an additional topological charge associated with the integer winding of a closed contour about the vortex axis.  We will explore numerical solutions with this winding in the next section.

\begin{figure}[ptb]
\centering
\includegraphics[width=0.8\linewidth]{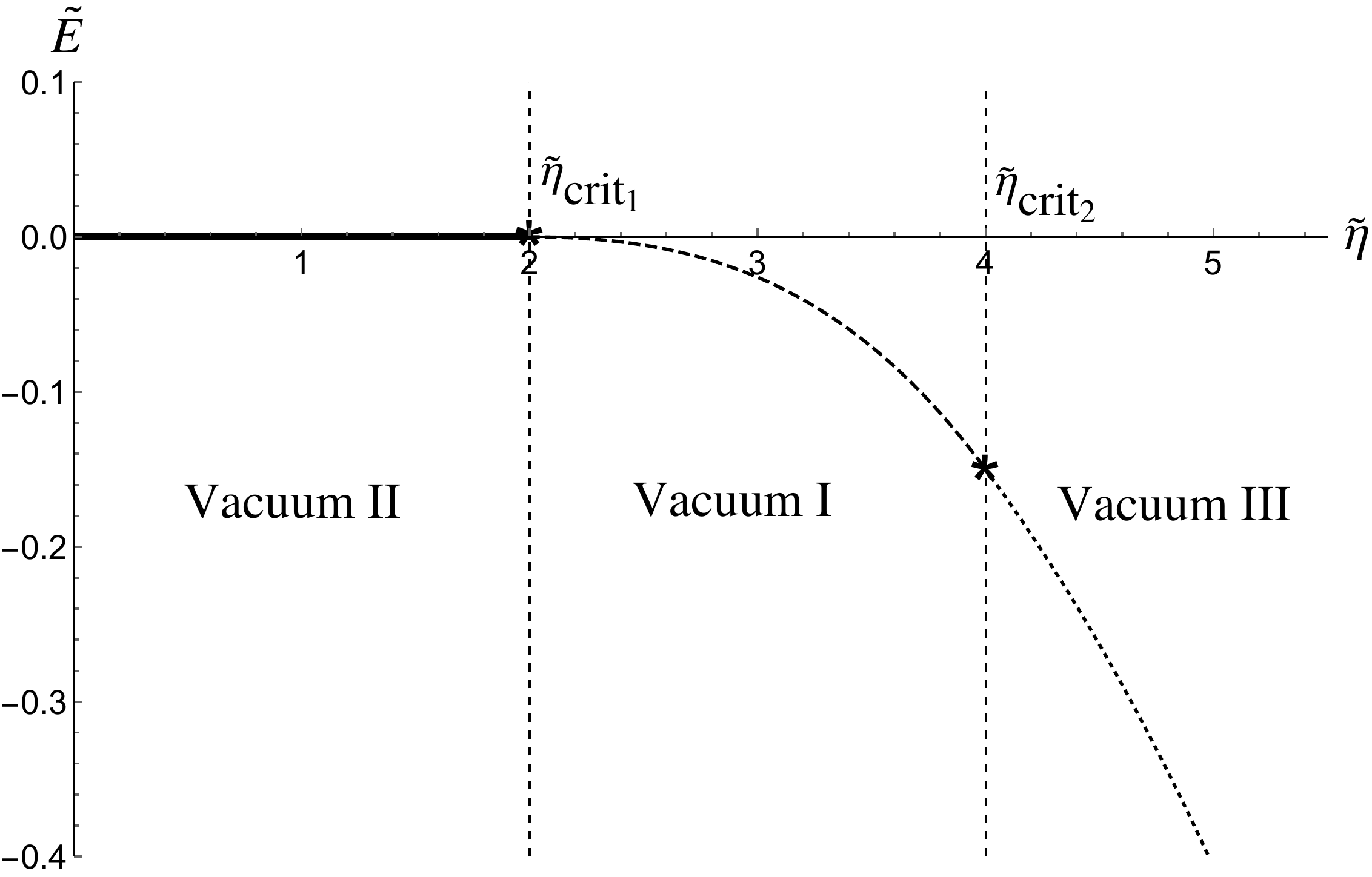}
\caption{$b=1,\; c=1.25,\; \beta=8$. Vacuum energy dependence on $\tilde{\eta}$ for $\beta(c-1)>bc$, the solid line corresponds to vacuum $II$, the dashed line to vacuum $I$ and the dotted line to vacuum $III$.}
\end{figure}

\begin{figure}[ptb]
\centering
\includegraphics[width=0.8\linewidth]{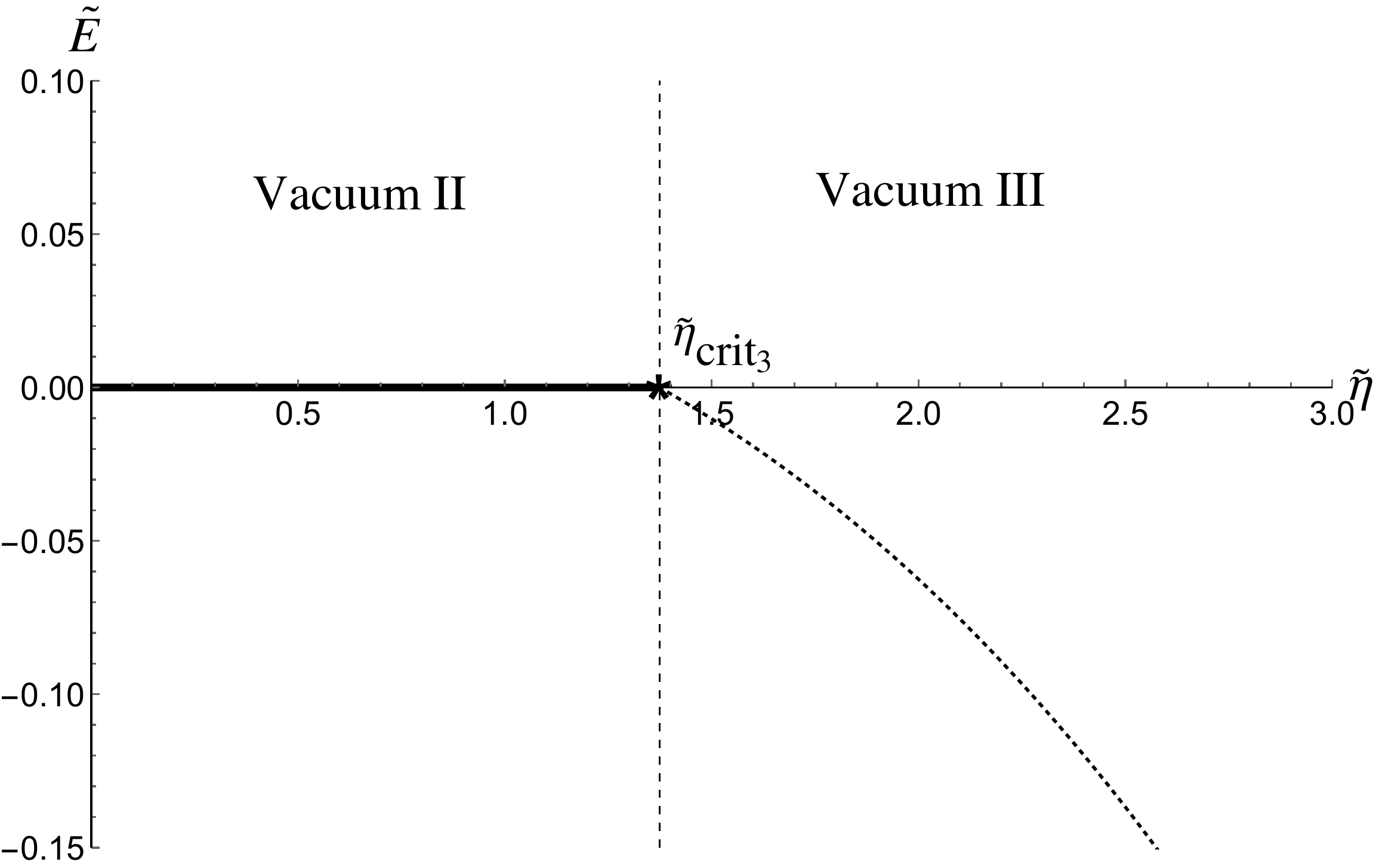}
\caption{$b=1,\; c=1.25,\; \beta=4$. Vacuum energy dependence on $\tilde{\eta}$ for $\beta(c-1)<bc$, the solid line corresponds to vacuum $II$ and the dotted line to vacuum $III$.}%
\end{figure}

\section{Winding solutions}

In \cite{Peterson:2014nma} the full set of eight coupled second order three dimensional equations was solved with a pseudo-spectral numerical procedure, additional details of this procedure are provided below (see \cite{Boyd:2001} for a detailed review of this method). The solution found, in which the spin field $\chi$ asymptotes to the vacuum at spatial infinity, describes a local $U(1)$ Abrikosov vortex with a cholesteric spiral inside its core. These vortices resemble the $w$ vortices appearing in superfluid $^3$He-B \cite{Salomaa:1987}. 

The degeneracy space $U(1)_{J''_z}$ allows for another class of solutions similar to the ``spin-mass-vortices" (see \cite{Kondo:1992}) with a global $U(1)_{J''_z}$ winding, characterized by a dependence of the $\chi_i$ field on the polar angle $\theta$ at infinity.  The $\theta$ dependence of $\chi_i$ at infinity is equivalent to translations of the vacuum state in the $\hat{k}$ direction.  The compactified translational group allows for the topological integer winding about a closed spatial contour. 

As a consequence of the global nature of these vortices the profile functions at large $\rho$ are IR divergent and require a regularization. Nonetheless, these solutions are important in the context of superfluid $^3$He \cite{Volovik:1977}. We are interested in solutions with spontaneous breaking of gauge symmetry, hence we only consider vacua $I$ and $II$ below.  

The geometry of our system is shown in Figure 3.  In order to find the $U(1)_{J''_z}$ winding solutions we switch to axially symmetric coordinates and use the ansatz
\bea
\chi_i &=& \frac{\mu}{\sqrt{2\beta}}(\chi_r(\rho,z),\tilde{\chi}_\theta(\rho,z),\chi_z(\rho,z)),\\
A_r &=&A_z=0\,,\quad A_\theta = 1-f(\rho)\,,
\nonumber\\[2mm]
\phi &=&  v\varphi(\rho)\, e^{i\theta}\,.
\eea
Then extremization of the energy density gives the following equations, 
\begin{align}
&\frac{1}{\rho}\partial_\rho\left(\rho\partial_\rho\chi_z\right)+\partial^2_z\chi_z-\tilde{\eta}\left(\partial_\rho\tilde{\chi}_\theta+\frac{\tilde{\chi}_\theta}{\rho}\right)-\frac{b}{c-1}(c\varphi^2 +\vec{\chi}^2 -1)\chi_z=0,\nn\\[4mm]
&\frac{1}{\rho}\partial_\rho\left(\rho\partial_\rho\tilde{\chi}_\theta\right)+\partial^2_z\tilde{\chi}_\theta-\frac{\tilde{\chi}_\theta}{\rho^2}-\tilde{\eta}(\partial_z\chi_r-\partial_\rho\chi_z)-\frac{b}{c-1}(c\varphi^2 + \vec{\chi}^2  -1)\tilde{\chi}_\theta=0,\nn\\[4mm]
&\frac{1}{\rho}\partial_\rho\left(\rho\partial_\rho\chi_r\right)+\partial^2_z\chi_r-\frac{\chi_r}{\rho^2}+\tilde{\eta}\partial_z\tilde{\chi}_\theta-\frac{b}{c-1}(c\varphi^2 +\vec{\chi}^2 -1)\chi_r=0,\nn\\[4mm]
&\frac{1}{\rho}\partial_\rho(\rho\partial_\rho\varphi)+\partial_z^2\varphi-\left[\frac{1}{2}(\varphi^2-1)+\frac{1}{\rho^2}f^2+\frac{b}{2\beta(c-1)}\vec{\chi}^2 \right]\varphi=0,\nn\\[4mm] 
&\partial_\rho^2f-\frac{1}{\rho}\partial_\rho f +\partial_z^2 f-a\varphi^2f=0.
\label{eom1}
\end{align}
We wish to solve the equations (\ref{eom1}) with the following boundary conditions 
\bea\label{boundary}
\varphi(0) =0 ,& \varphi(\infty)=\varphi_0,\nn\\[3mm]
f(0)=1, & f(\infty)=0,\nn\\[3mm]
\chi_z'(0,z)=0,& \chi_z(\infty,z)=0,\nn\\[3mm]
\chi_r(0, z)=0, & \chi_r(\infty, z)=\chi_0 \cos\left(\frac{\eta}{2}z\right),\nn\\[3mm]
\tilde{\chi}_\theta(0, z)=0, & \tilde{\chi}_\theta(\infty, z)=\chi_0 \sin\left(\frac{\eta}{2}z\right),
\eea
where $\chi_0$ and $\varphi$ are the same as the previous section and we demand continuity in the $z$ direction.
In terms of the cartesian vector this boundary condition takes the form
\be
\chi_i = \chi_0 \big \{\cos(kz+\theta),\; \sin(kz+\theta),\;0\big\}
\ee
which is not to be confused with the $U(1)$ vortex solution of \cite{Peterson:2014nma}. In fact, it is only because of this subtle difference between asymptotic conditions that the full set of three dimensional equations (\ref{eomm}) becomes two dimensional. This greatly simplifies the numerical procedure.

\begin{figure}[ptb]
\centering
\includegraphics[width=0.8\linewidth]{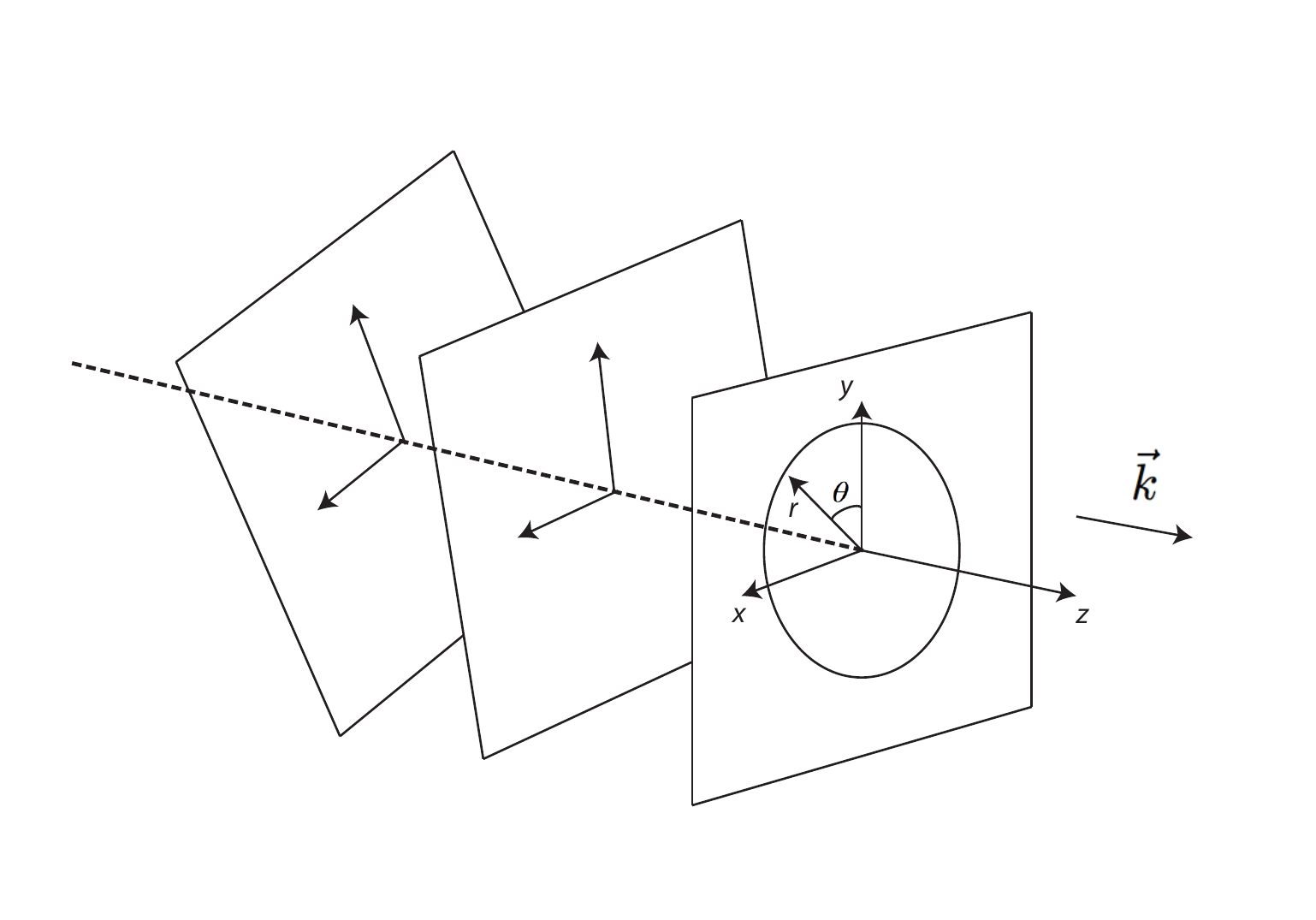}
\caption{Geometric set-up of the problem. The vortex axis, in the $z$ direction, is parallel to the wave-vector $\vec{k}$ which is normal to the cholesteric
planes.
}
\end{figure}

\subsection{Outline of numerical procedure}

The reduction of the spatial dependence of the coupled system (\ref{eom1}) from three to two coordinates greatly simplifies the numerical analysis by reducing the computing power needed.  However, for precise numerical results, two dimensional finite difference methods are still impractical when efficiency and precision are the goals.  To achieve these goals we chose instead to use a pseudospectral method on a collocated grid.  In this section we will give a brief outline of our procedure.  A detailed and very readable description of these methods can be found in \cite{Boyd:2001}.

For the $z$ dependence of the functions defined above, we choose to expand in Fourier modes and evaluate the functions at the discrete points $z_m = 2\pi m/kM$, where $M$ is the total number of Fourier modes used in the decomposition.  To decompose the $r$ dependence in the functions above, we first compactify the radial direction using the transformation:
\begin{equation}
r = \frac{Lx}{\sqrt{1-x^2}}, \; \; x \in [0,1].
\end{equation}
The parameter $L$ determines the distribution of grid points.  For this procedure we selected $L=2$ to achieve the desired density of grid points near the origin.
The radial dependence of the functions is then expanded in rational Chebyshev polynomials $T_n(x(r))$ on a collocated Radau grid:
\begin{equation}
x_p = \sin{\frac{\pi(2p-1)}{4N-2}}, \;\; n \in [1,N],
\end{equation} 
where $N$ is the total number of Chebyshev polynomials used in the decomposition.  The Radau grid allows for boundary conditions to be imposed at spatial infinity with no conditions required at the origin.  For functions required to vanish at the origin we expand in odd Chebyshevs.

The pseudospectral decomposition allows the equations (\ref{eom1}) to be written as a set of algebraic equations to be solved for the coefficients in the expansion.  For the five functions to be solved for above, this amounts to solving a coupled system of $5\times N\times M$ polynomial equations.  To approximate a solution the system is linearized and solved recursively.  Each iteration of the procedure is repeated until convergence is achieved.

\subsection{Vacuum I}

Let us first consider solutions in which $\chi_0 \neq0$. In this case the vacuum of the system is vacuum I, given by (\ref{E1}). Clearly in this case the solutions with boundary conditions (\ref{boundary}) don't approach the vacuum at infinity and are logarithmically divergent in the IR.  These solutions require an IR regularization. We restrict our solutions to a finite radial distance $L$. The winding of these spin-vortices ensures their topological stability.  It should be pointed out that near the vortex core a non-zero value of $\chi_z$ appears due to the variation of $\chi_\theta$ in the radial direction.  

For parameter values $a=1, \; b=1, \; c = 1.25, \; \beta = 8,$ and $\eta = 2.5$ we obtain the solutions shown in Figures 4-6.

\begin{figure}[ptb]
\centering
\includegraphics[width=0.8\linewidth]{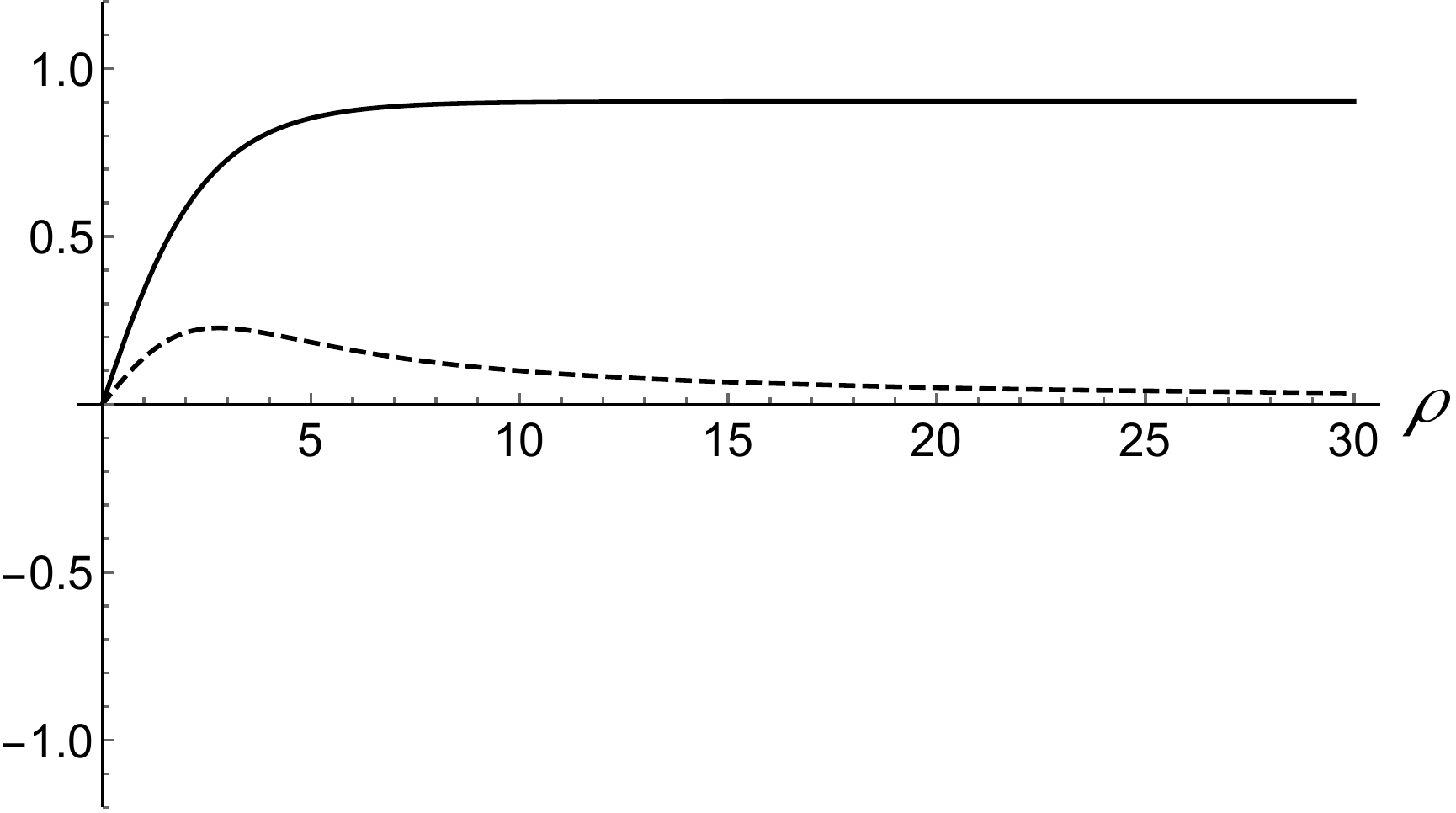}
\caption{The graph shows the solutions for $\phi(r)$ (solid) and $A_\theta (r)$ (dashed) at $z=0$, for $\eta = 2.5$.  These solutions have negligable $z$ dependence.}%
\end{figure}

\begin{figure}
\hspace*{-2.4cm}
\begin{subfigure}{5cm}
\centering
\includegraphics[width=1.5\linewidth]{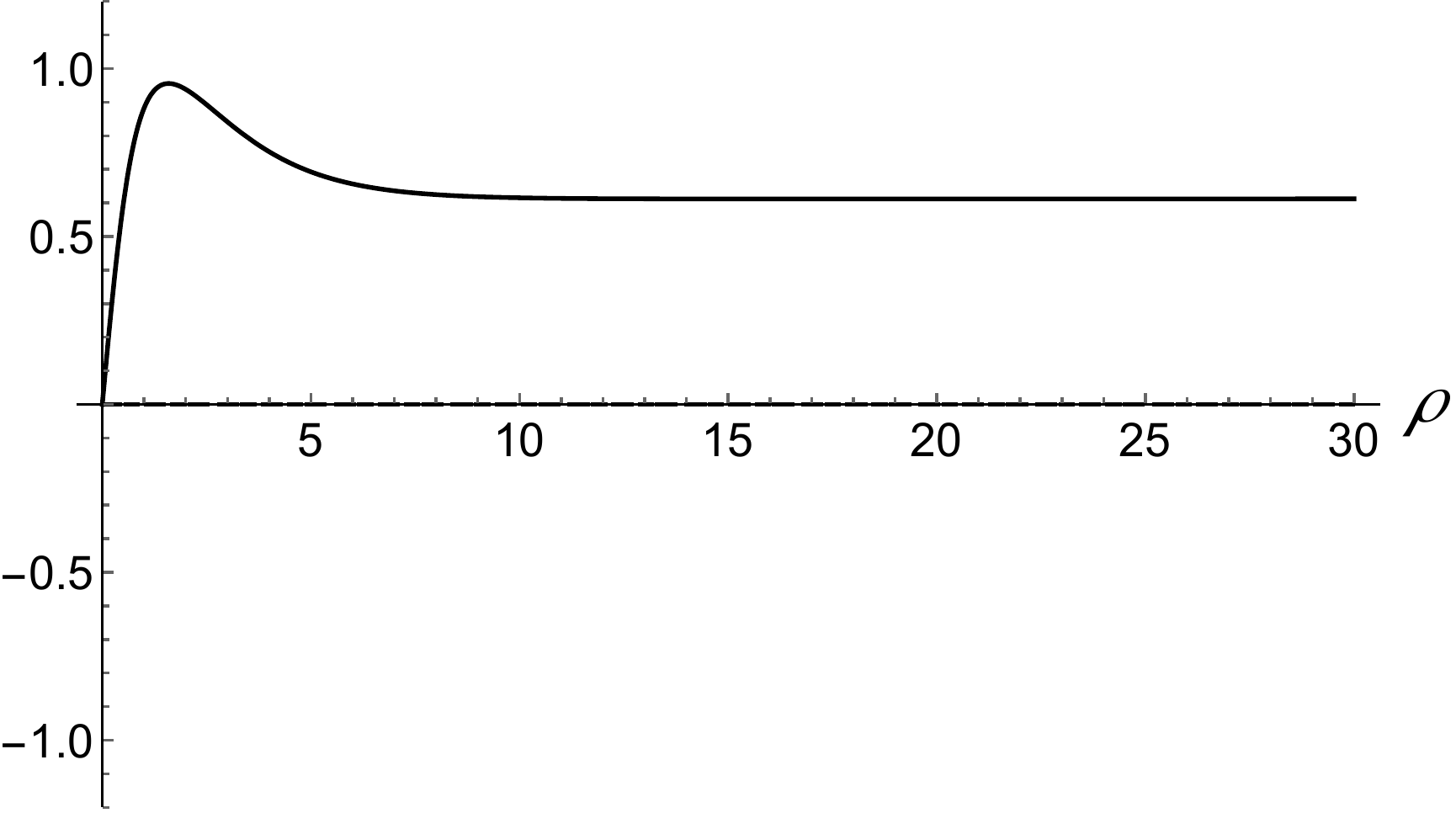}
\caption{The field $\vec{\chi}$, at $kz=0$.}
\end{subfigure}%
\quad\quad\quad\quad\quad\;\;\;\;\;\;\;\;\;
\begin{subfigure}{5cm}
\centering
\includegraphics[width=1.5\linewidth]{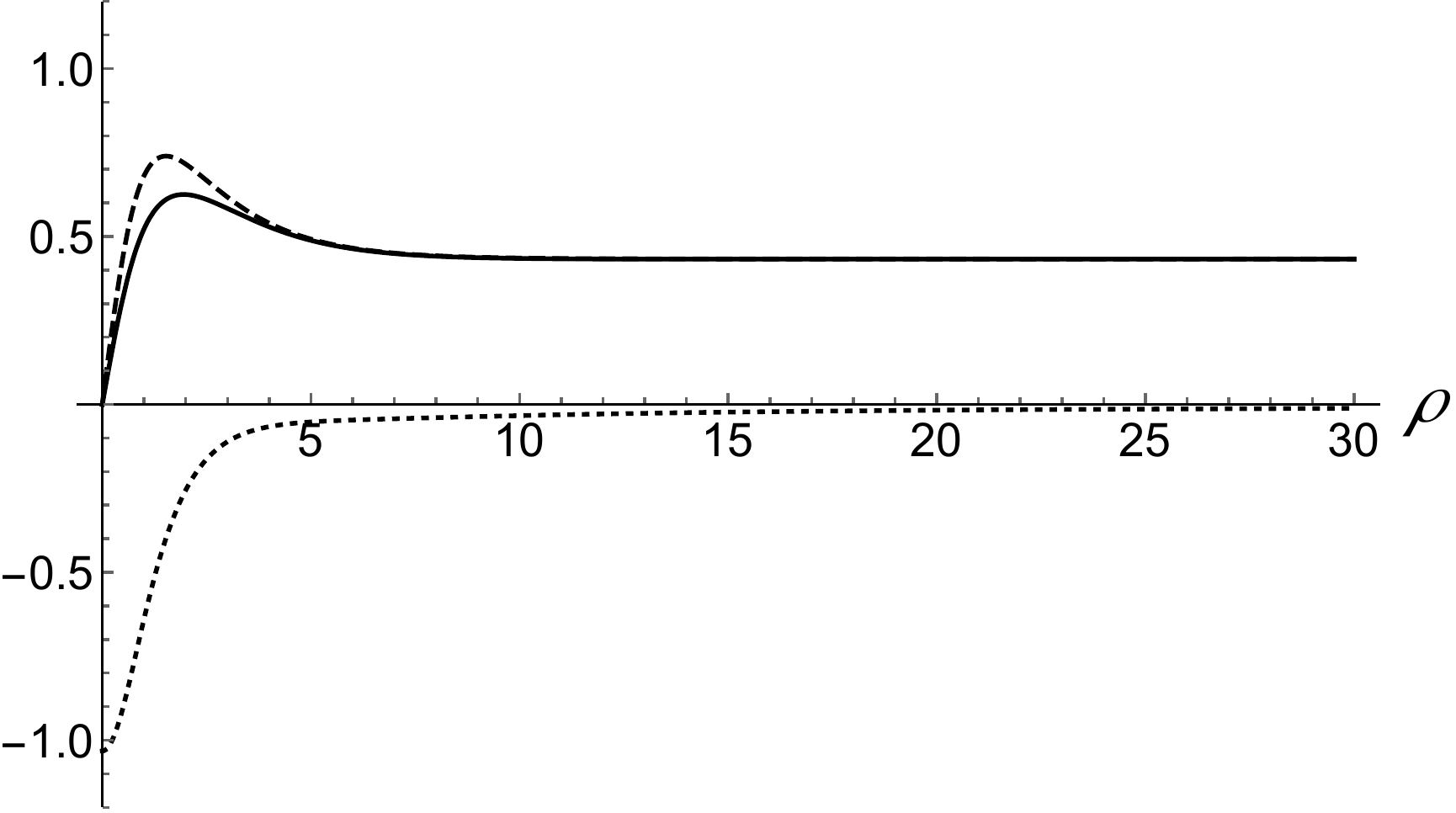}
\centering
\caption{The field $\vec{\chi}$, at $kz=\pi/4.$}
\end{subfigure}\\
\\
\\
\centering
\hspace*{-2.4cm}
\begin{subfigure}{5cm}
\centering
\includegraphics[width=1.5\linewidth]{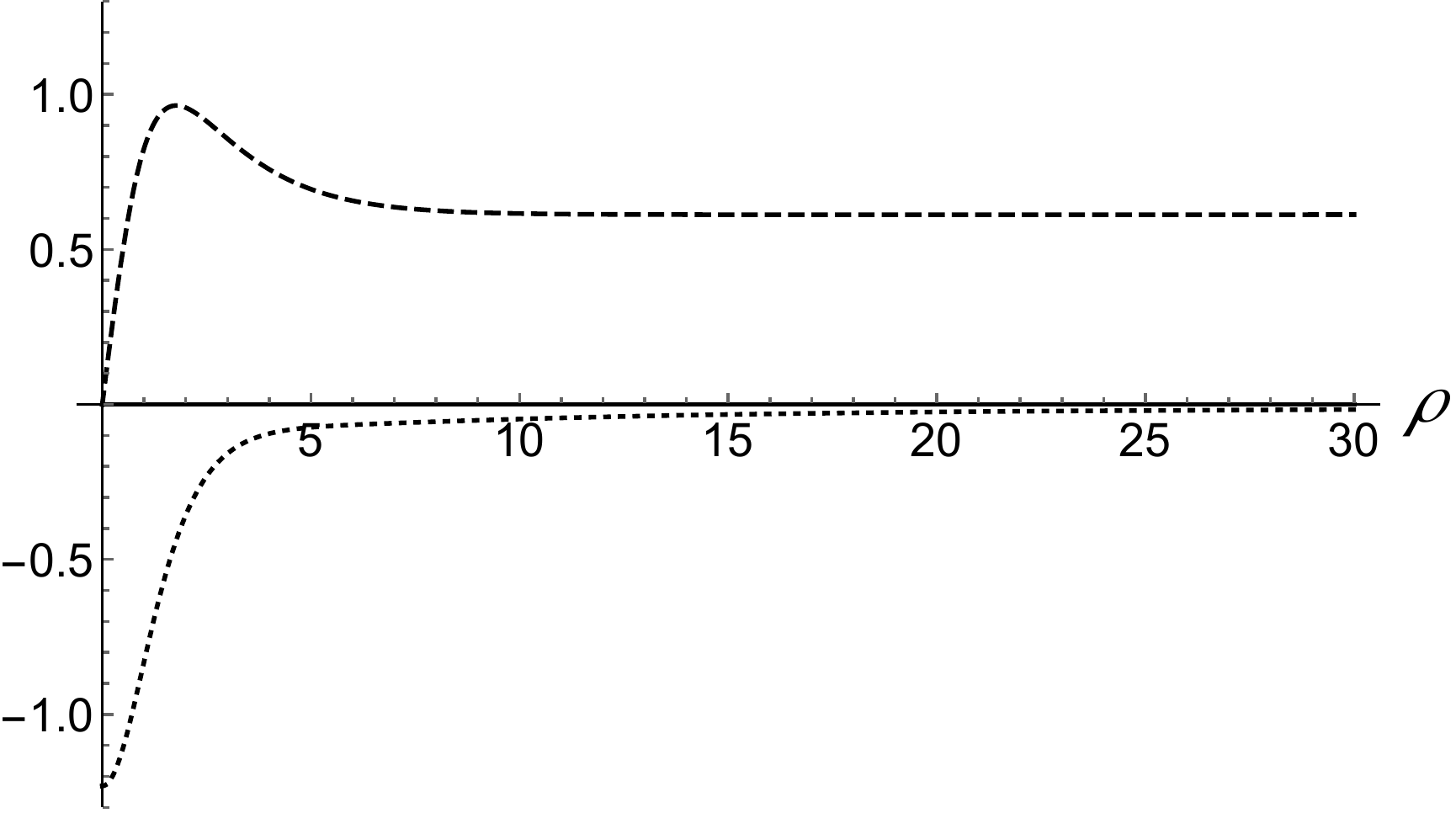}
\caption{The field $\vec{\chi}$, at $kz=\pi/2.$}
\end{subfigure}%
\quad\quad\quad\quad\quad\;\;\;\;\;\;\;\;\;
\begin{subfigure}{5cm}
\centering
\includegraphics[width=1.5\linewidth]{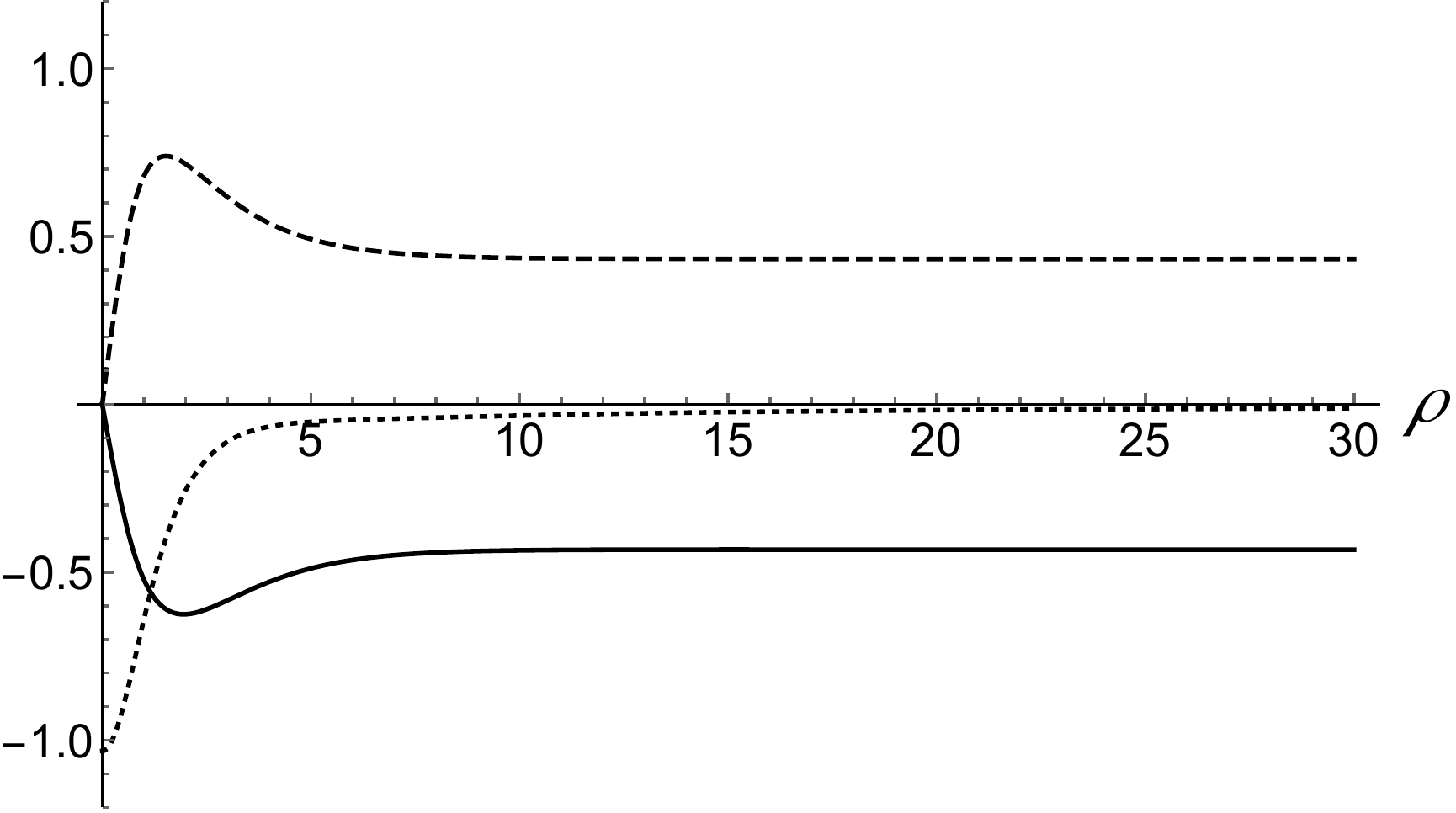}
\centering
\caption{The field $\vec{\chi}$, at $kz=3\pi/4.$}
\end{subfigure}\\
\caption{The graphs show the radial dependence of $\chi_r$ (solid), $\chi_\theta$ (dashed), and $\chi_z$ (dotted) at various values of $kz$.}
\end{figure}

\begin{figure}
\hspace*{-2.4cm}
\begin{subfigure}{5cm}
\centering
\includegraphics[width=1.5\linewidth]{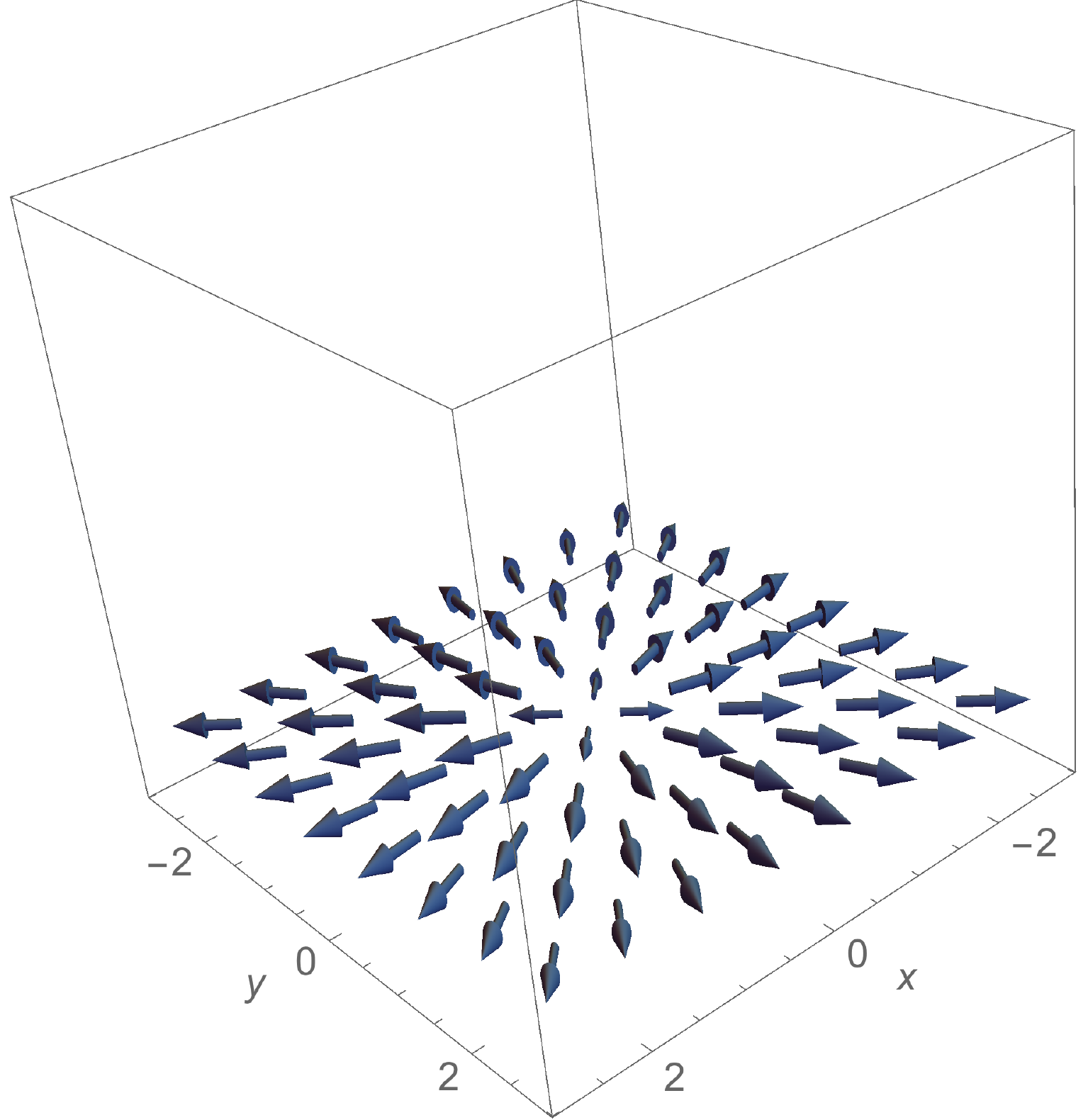}
\caption{The field $\vec{\chi}$, at $kz=0$.}
\end{subfigure}%
\quad\quad\quad\quad\quad\;\;\;\;\;\;\;\;\;
\begin{subfigure}{5cm}
\centering
\includegraphics[width=1.5\linewidth]{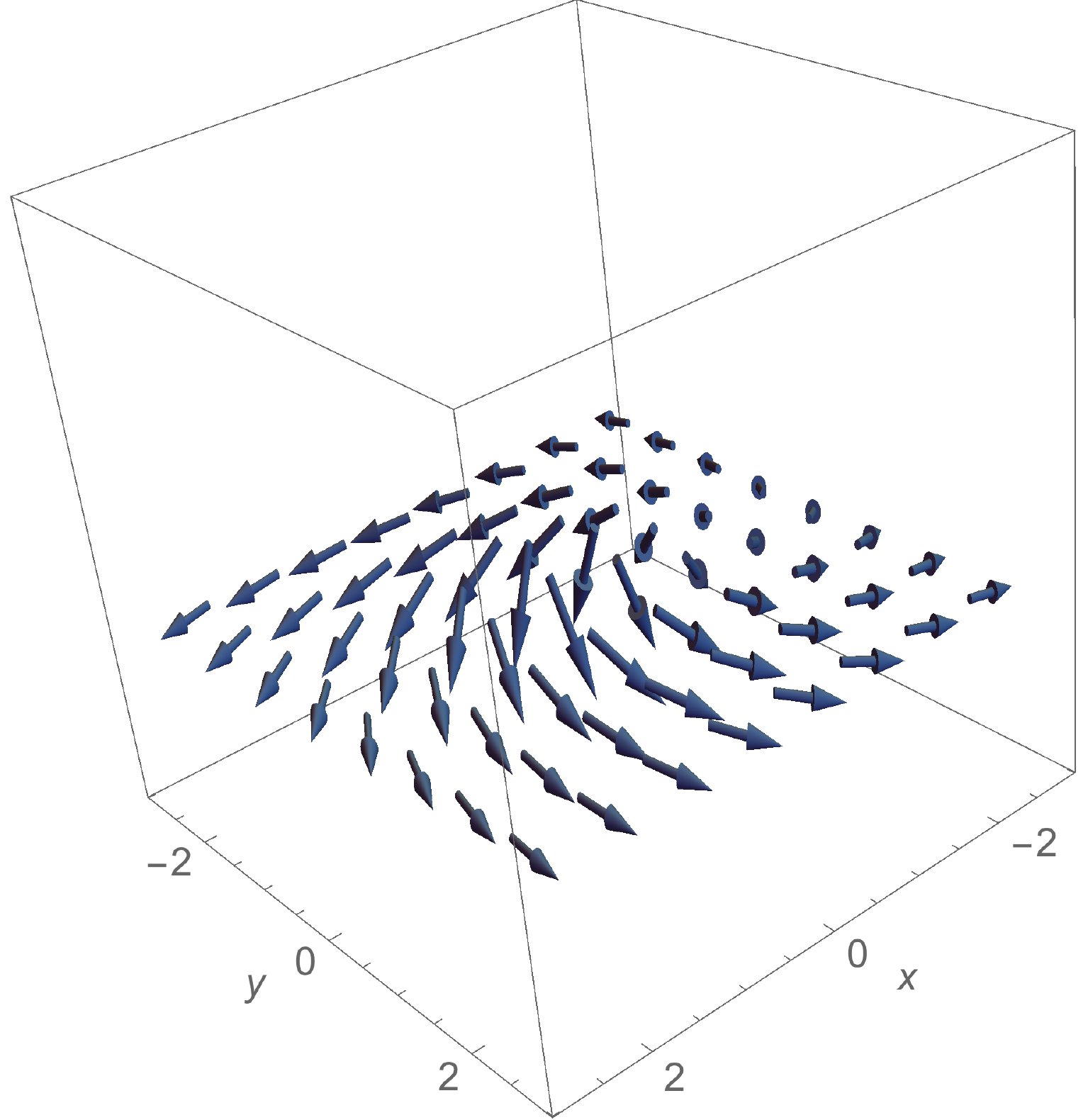}
\centering
\caption{The field $\vec{\chi}$, at $kz=\pi/4.$}
\end{subfigure}\\
\\
\\
\centering
\hspace*{-2.4cm}
\begin{subfigure}{5cm}
\centering
\includegraphics[width=1.5\linewidth]{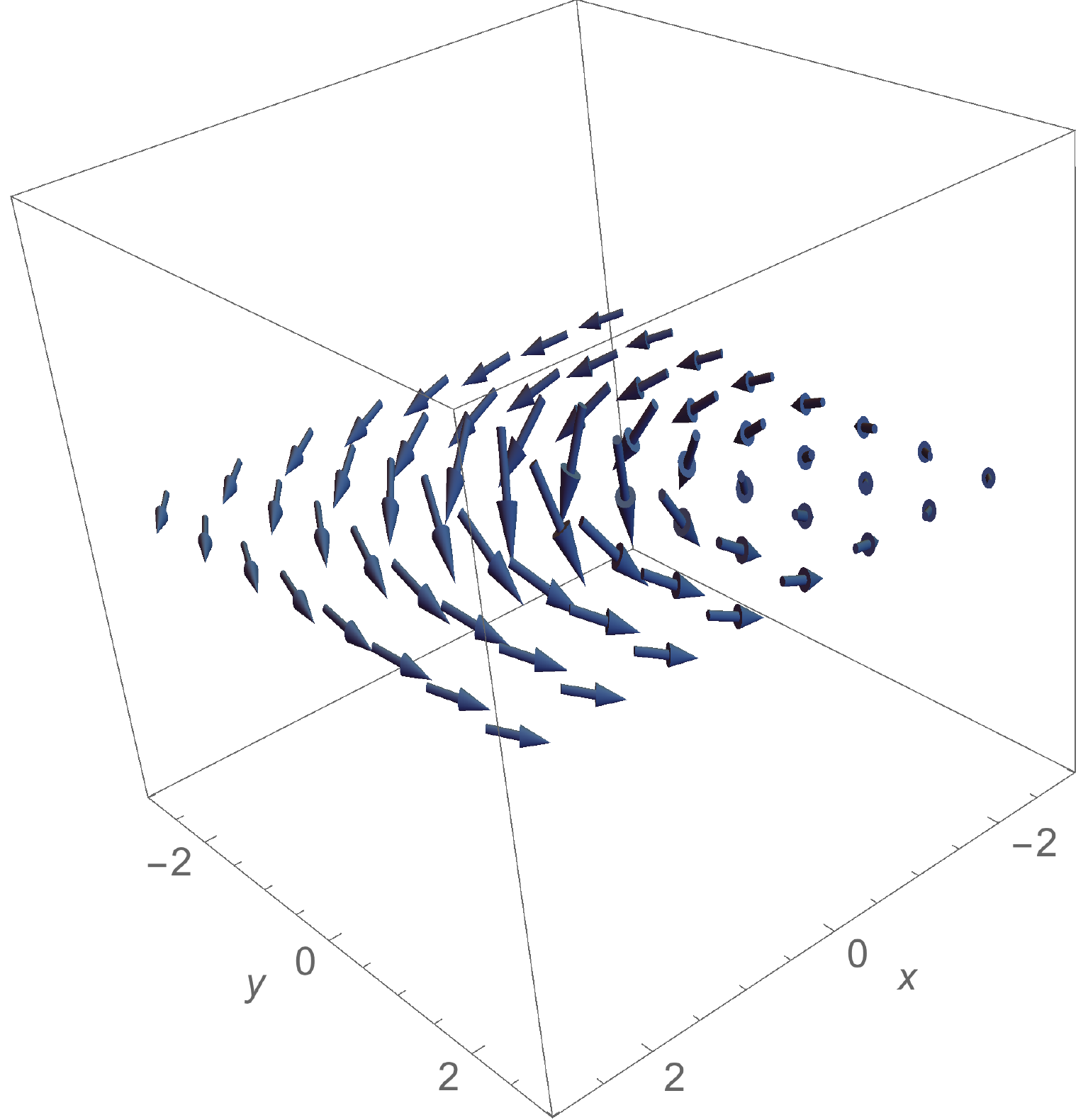}
\caption{The field $\vec{\chi}$, at $kz=\pi/2.$}
\end{subfigure}%
\quad\quad\quad\quad\quad\;\;\;\;\;\;\;\;\;
\begin{subfigure}{5cm}
\centering
\includegraphics[width=1.5\linewidth]{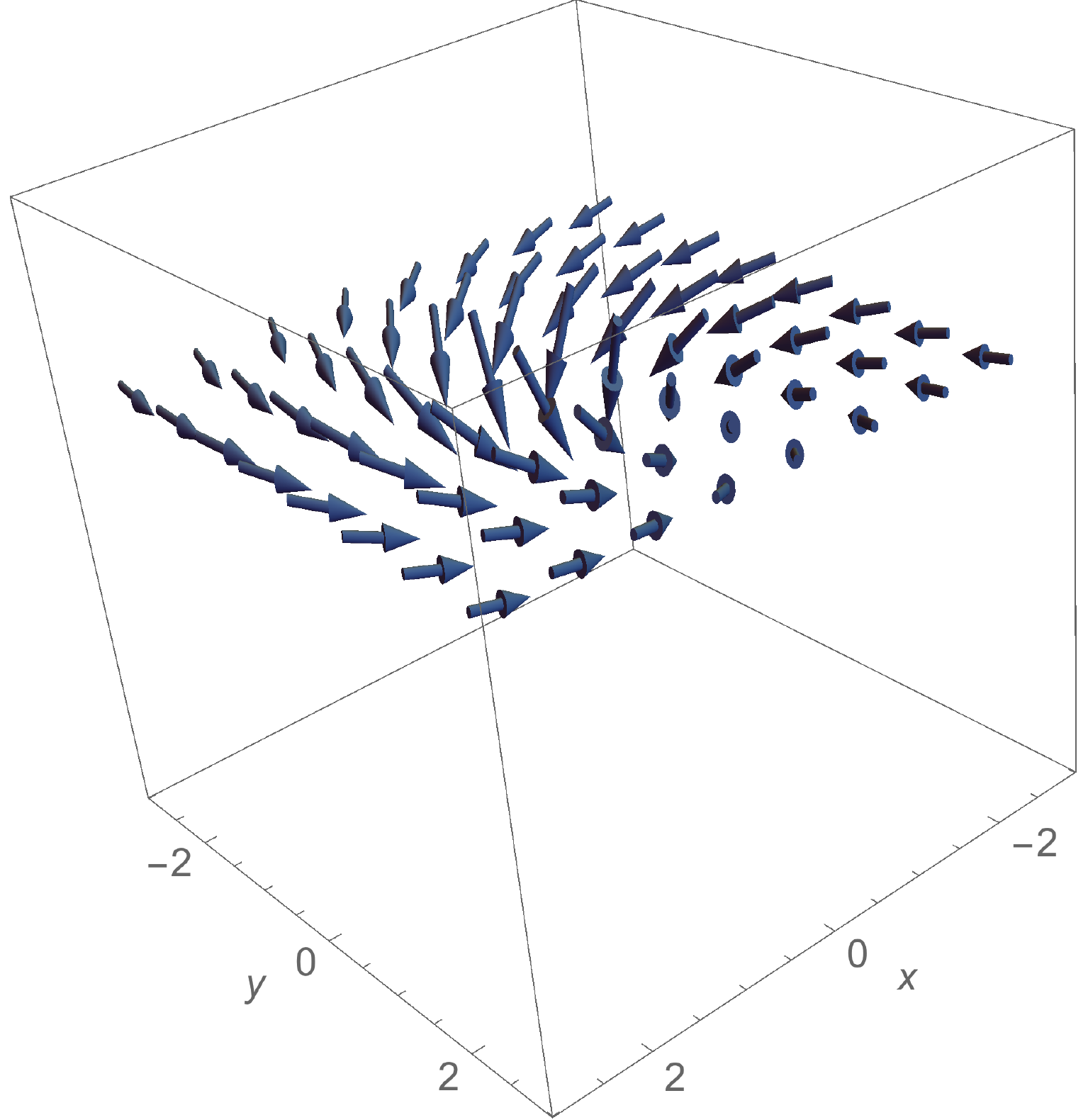}
\centering
\caption{The field $\vec{\chi}$, at $kz=3\pi/4.$}
\end{subfigure}\\
\caption{Shown are the vector fields $\vec{\chi}$ at various values of the coordinate $z \in [0,3\pi/4k]$ for $\eta = 2.5$.  We note that as the vortex core is approached a non-zero value of $\chi_z$ appears.  This is expected from the equations of motion when $\chi_\theta$ has an $r$-dependence near the core.}
\end{figure}

\subsection{Vacuum II}

When $\chi_0 =0$ the vacuum has the structure of vacuum II (\ref{Vac2}). In this case, the asymptotics of the solutions (\ref{boundary}) have no cholesteric structure and thus we expect finite energy solutions. However, the winding nature of these solutions is lost.  These solutions lack the topological stability of the previous case in vacuum I.  They are instead metastable solutions whose topological stability is only supported to finite radial distance from the core.  Presumably, these solutions will decay to the zero winding state through quantum tunneling via bubble nucleation.  We leave an investigation of the tunneling rate to another project.  Figures 7-9 illustrate these solutions for $a=1, \; b=1, \; c = 1.25, \; \beta = 8,$ and $\eta = 2.2$.

\begin{figure}[ptb]
\centering
\includegraphics[width=0.8\linewidth]{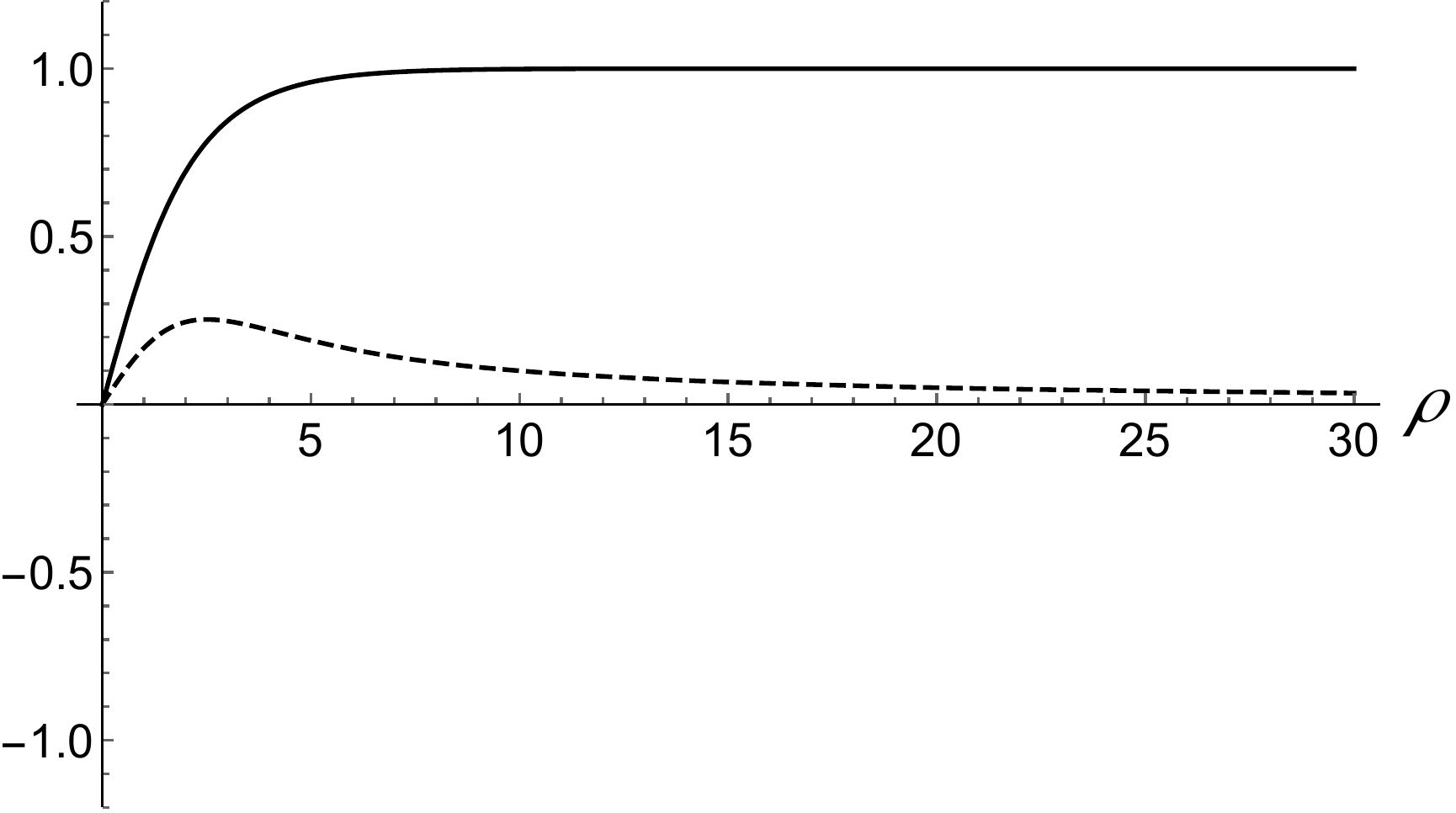}
\caption{The graph shows the solutions for $\phi(r)$ (solid) and $A_\theta (r)$ (dashed) at $z=0$, for $\eta = 2.2$.  These solutions have negligable $z$ dependence.}%
\end{figure}

\begin{figure}
\hspace*{-2.4cm}
\begin{subfigure}{5cm}
\centering
\includegraphics[width=1.5\linewidth]{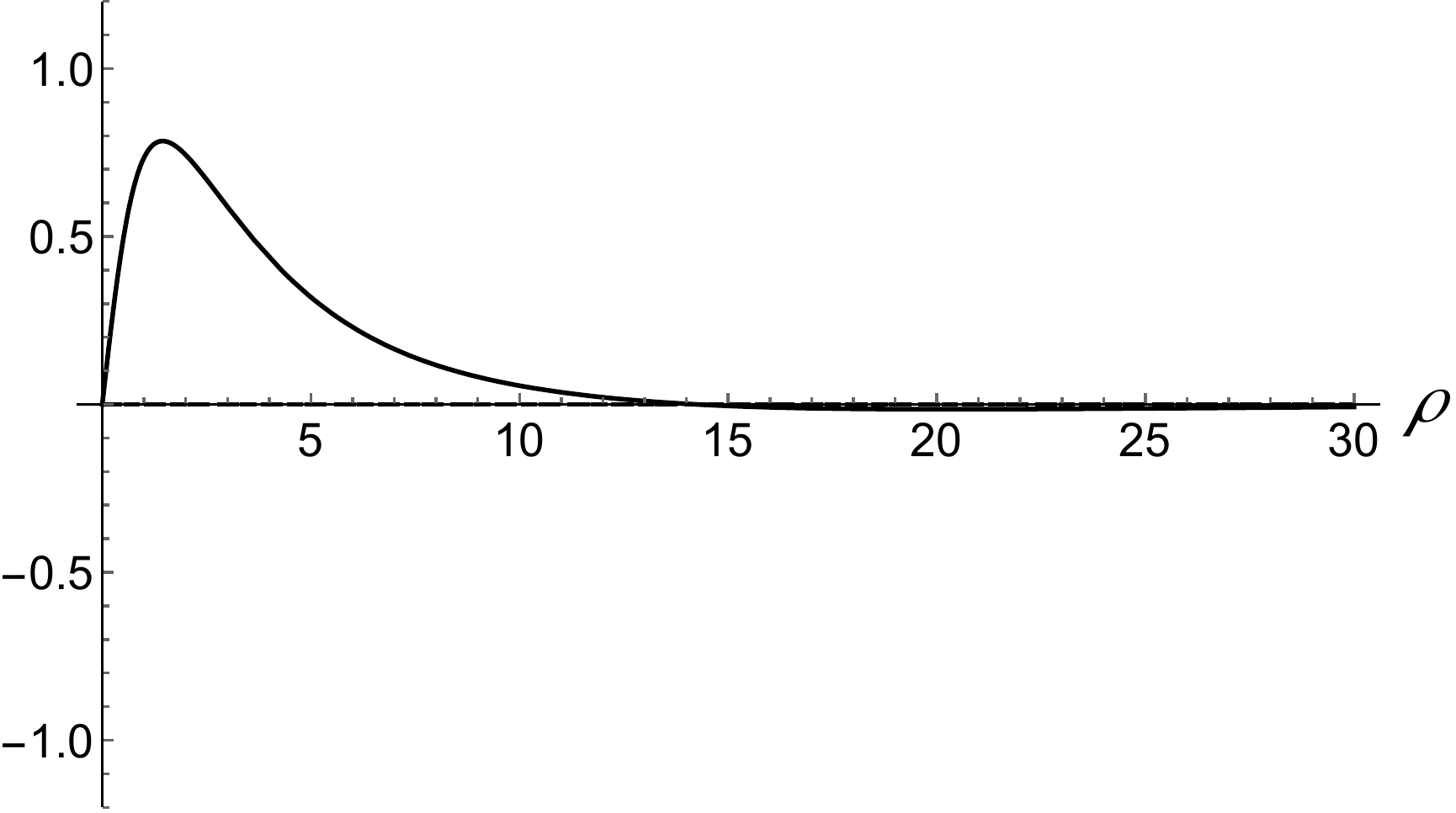}
\caption{The field $\vec{\chi}$, at $kz=0$.}
\end{subfigure}%
\quad\quad\quad\quad\quad\;\;\;\;\;\;\;\;\;
\begin{subfigure}{5cm}
\centering
\includegraphics[width=1.5\linewidth]{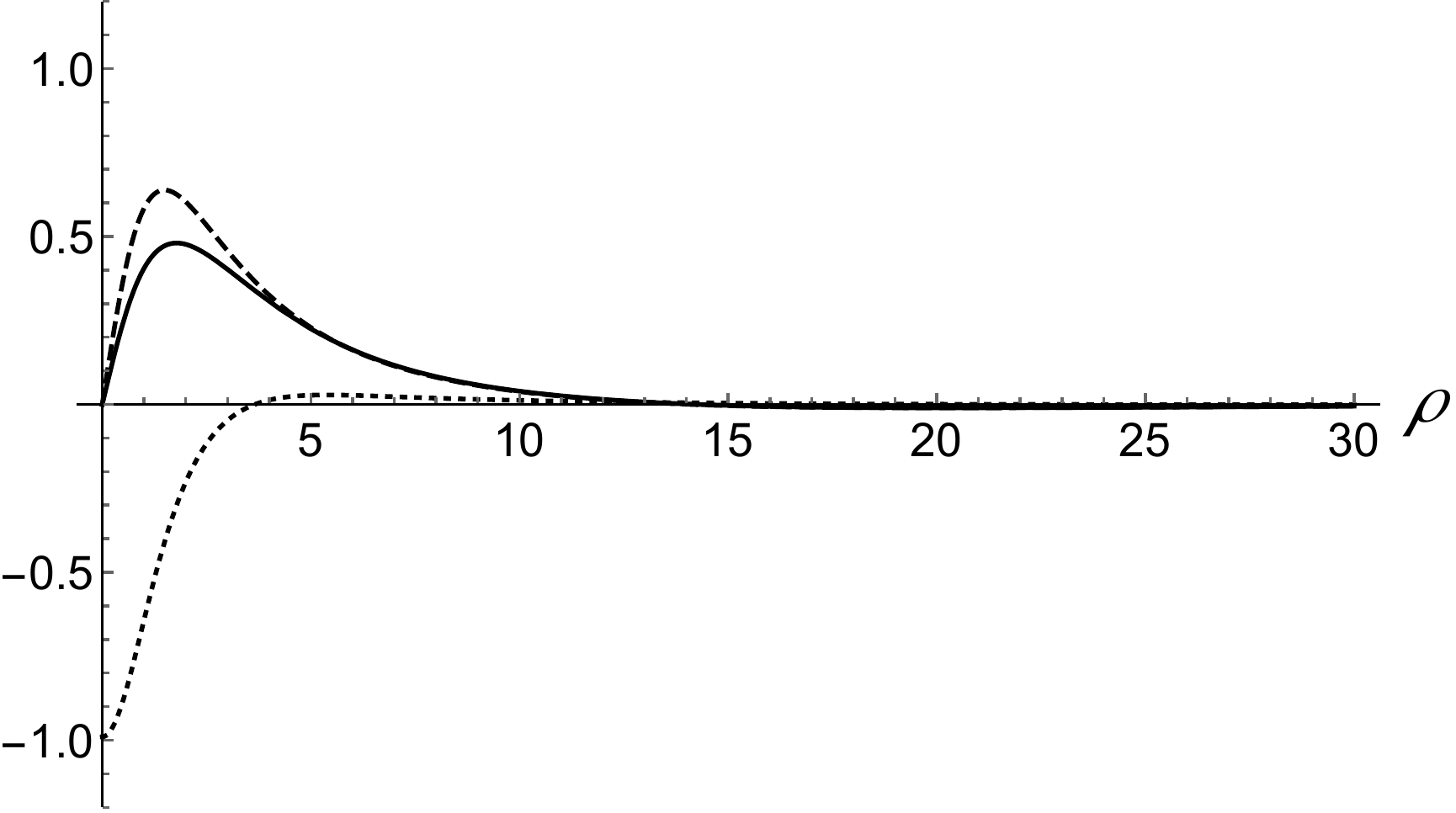}
\centering
\caption{The field $\vec{\chi}$, at $kz=\pi/4.$}
\end{subfigure}\\
\\
\\
\centering
\hspace*{-2.4cm}
\begin{subfigure}{5cm}
\centering
\includegraphics[width=1.5\linewidth]{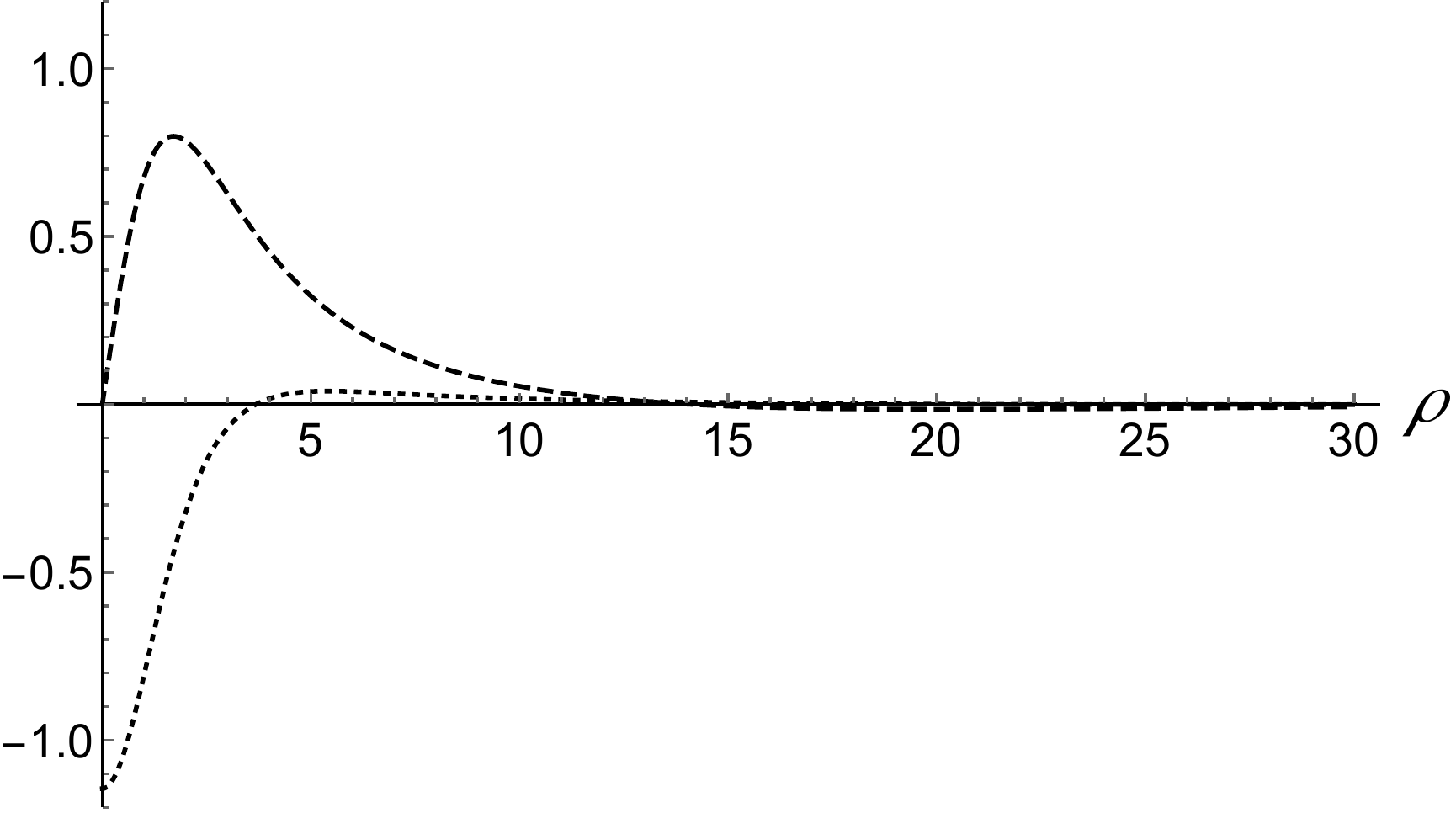}
\caption{The field $\vec{\chi}$, at $kz=\pi/2.$}
\end{subfigure}%
\quad\quad\quad\quad\quad\;\;\;\;\;\;\;\;\;
\begin{subfigure}{5cm}
\centering
\includegraphics[width=1.5\linewidth]{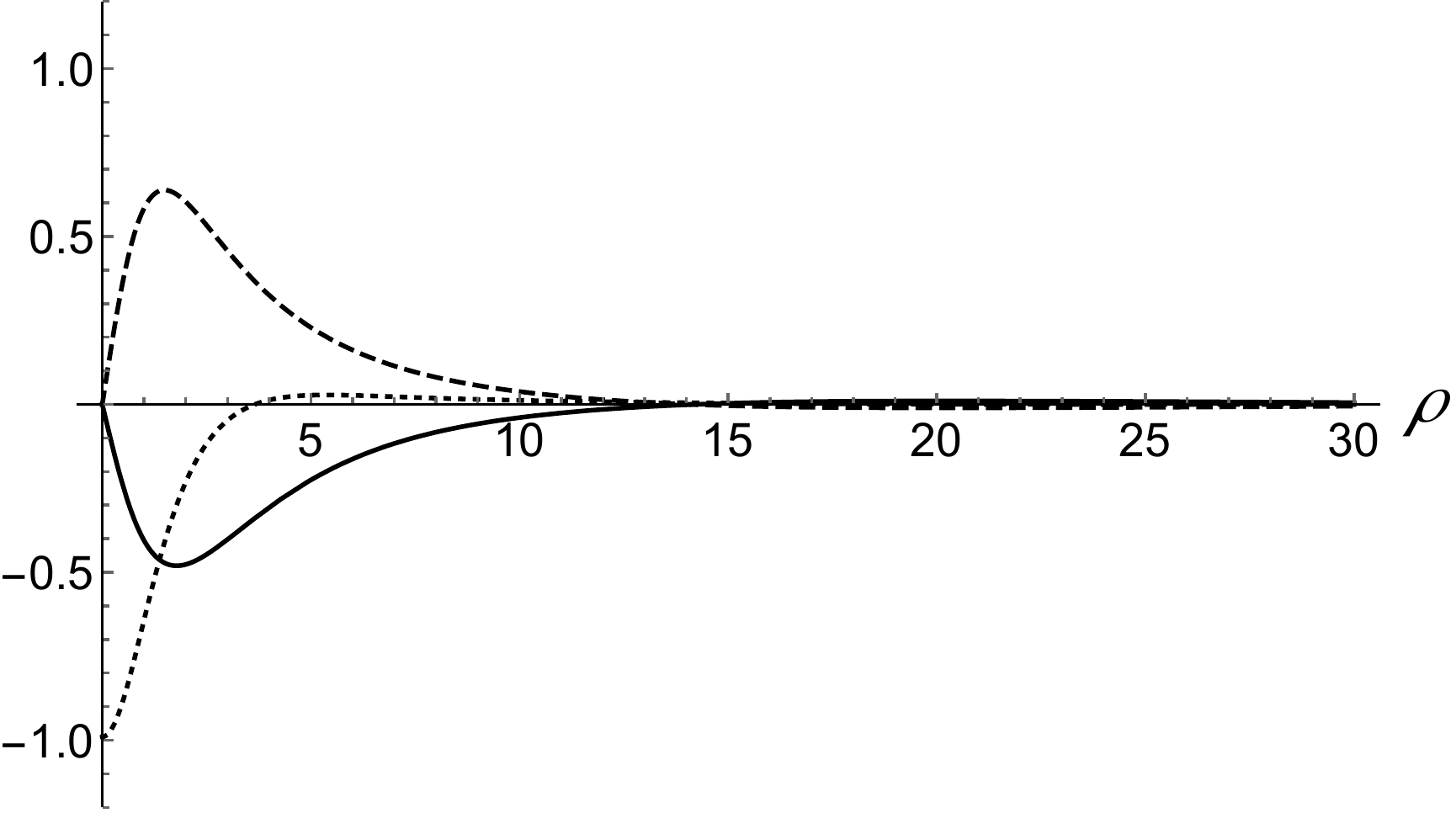}
\centering
\caption{The field $\vec{\chi}$, at $kz=3\pi/4.$}
\end{subfigure}\\
\caption{The graphs show the radial dependence of $\chi_r$ (solid), $\chi_\theta$ (dashed), and $\chi_z$ (dotted) at various values of $kz$.}
\end{figure}

\begin{figure}
\hspace*{-2.4cm}
\begin{subfigure}{5cm}
\centering
\includegraphics[width=1.5\linewidth]{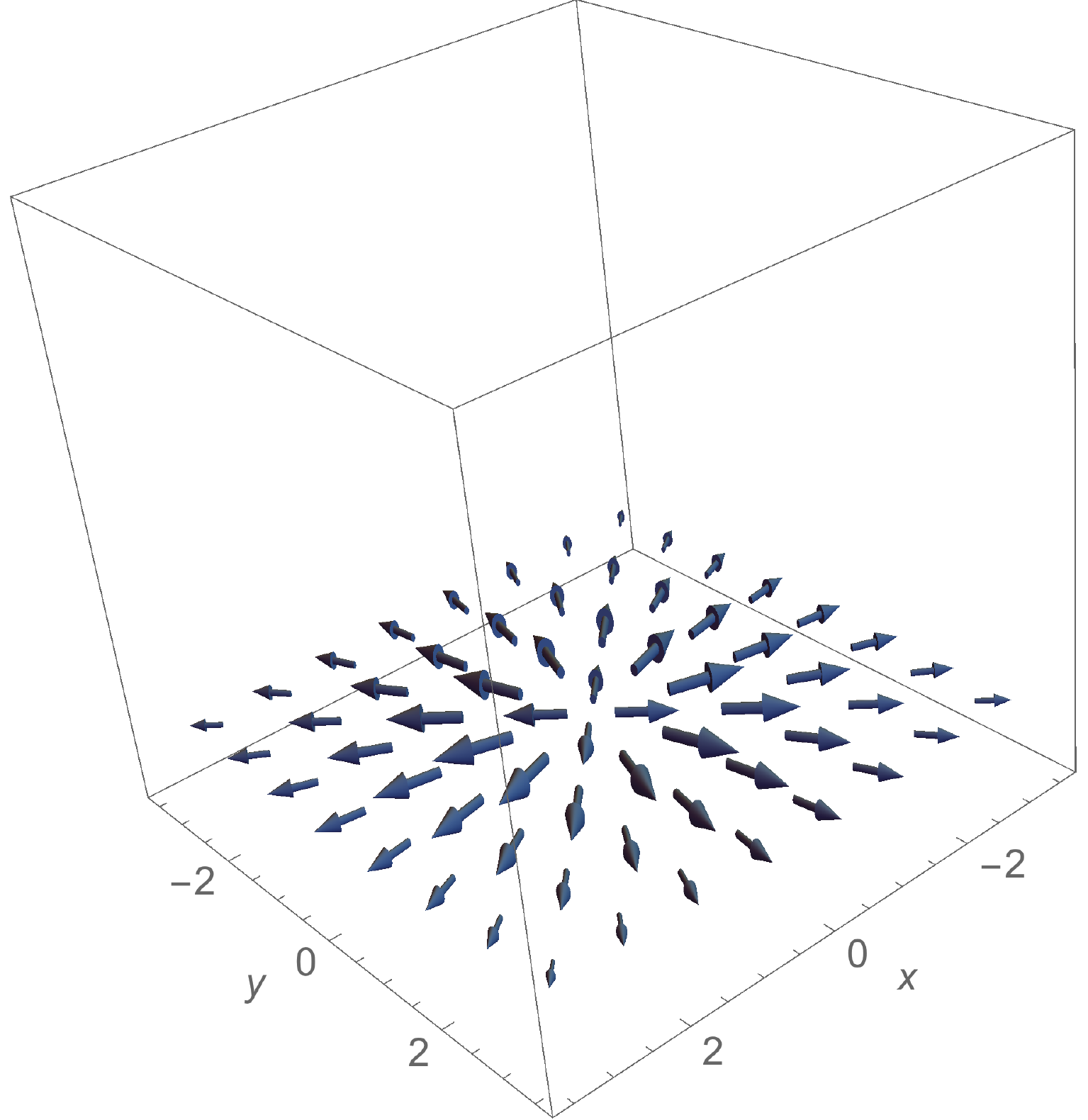}
\caption{The field $\vec{\chi}$, at $kz=0$.}
\end{subfigure}%
\quad\quad\quad\quad\quad\;\;\;\;\;\;\;\;\;
\begin{subfigure}{5cm}
\centering
\includegraphics[width=1.5\linewidth]{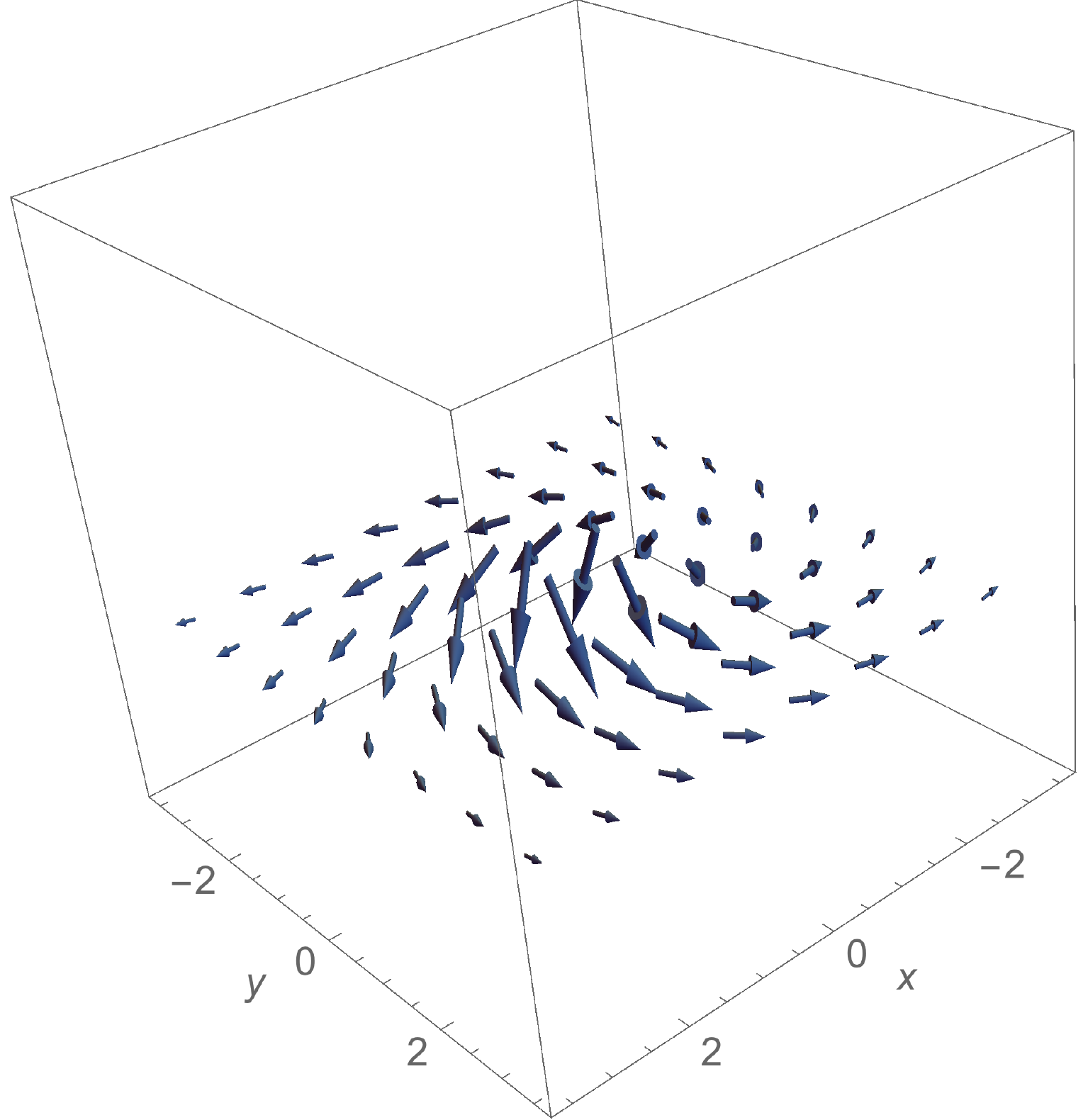}
\centering
\caption{The field $\vec{\chi}$, at $kz=\pi/4.$}
\end{subfigure}\\
\\
\\
\centering
\hspace*{-2.4cm}
\begin{subfigure}{5cm}
\centering
\includegraphics[width=1.5\linewidth]{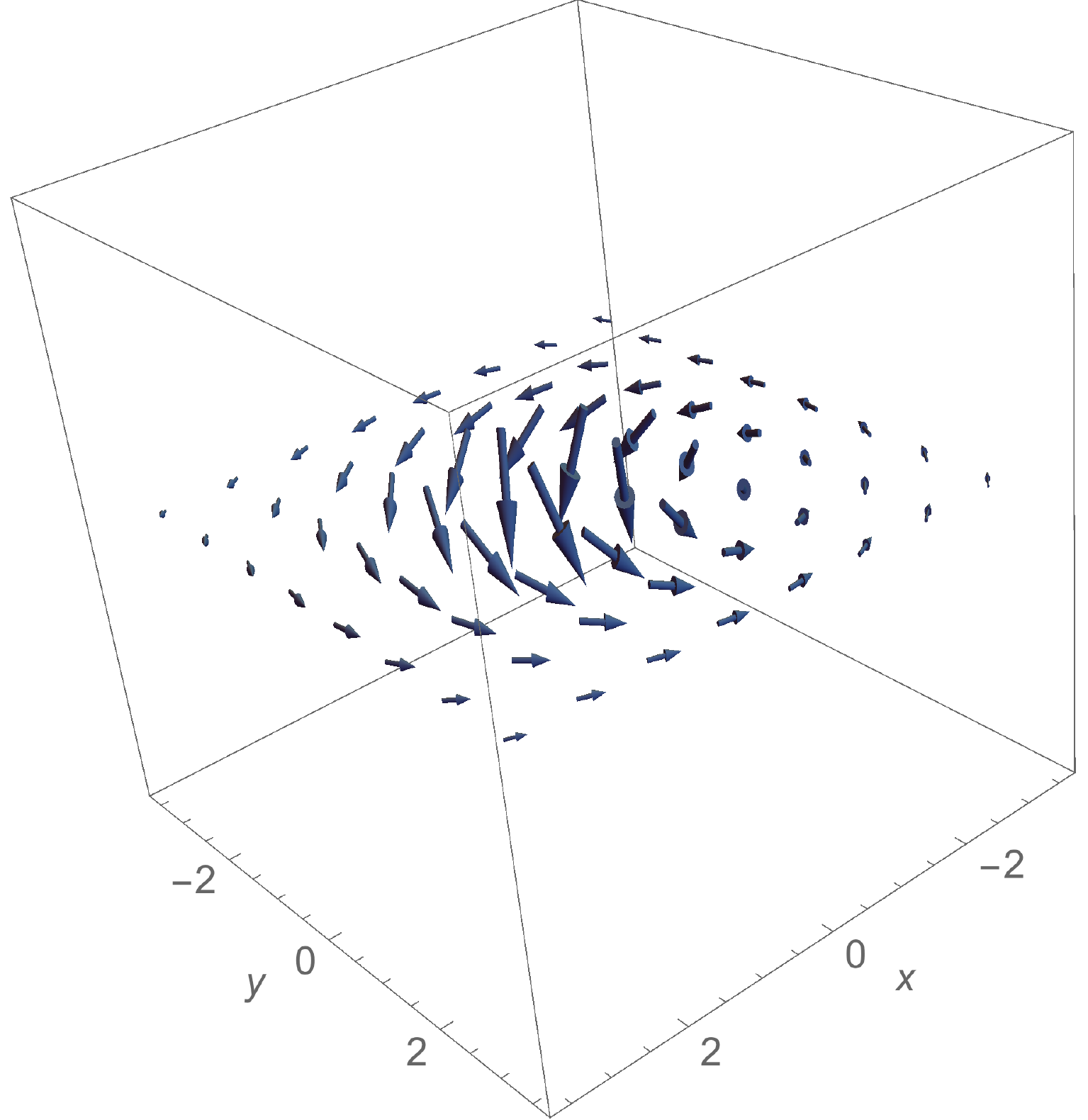}
\caption{The field $\vec{\chi}$, at $kz=\pi/2.$}
\end{subfigure}%
\quad\quad\quad\quad\quad\;\;\;\;\;\;\;\;\;
\begin{subfigure}{5cm}
\centering
\includegraphics[width=1.5\linewidth]{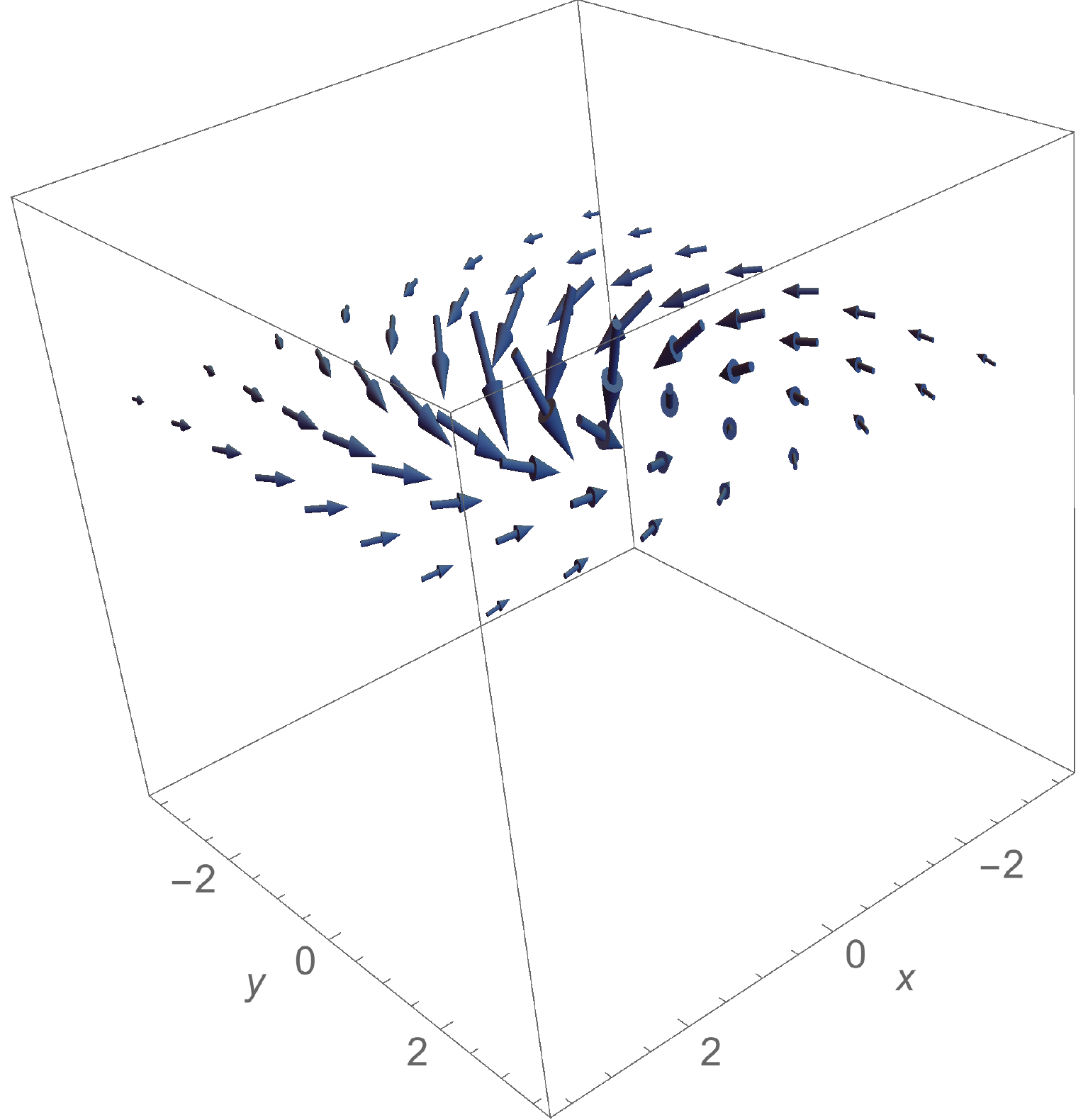}
\centering
\caption{The field $\vec{\chi}$, at $kz=3\pi/4.$}
\end{subfigure}\\
\caption{Shown are the vector fields $\vec{\chi}$ at various values of the coordinate $z \in [0,3\pi/4k]$ for the critical point $\eta = 2$.  We note that as the vortex core is approached a non-zero value of $\chi_z$ appears.  This is expected from the equations of motion when $\chi_\theta$ has an $r$-dependence near the core.}
\end{figure}

\section{Low energy theory of vortex moduli}

To complete the analysis of the spin vortices in vacuums $I$ and $II$, we discuss the low energy dynamics of their excitations localized on the vortex core.  In \cite{Peterson:2014nma} the low energy expansion was done at least for small $\eta$ for both the translational (Kelvin) and orientational moduli appearing in vacuum $I$ and $II$.  Here we will do an analogous treatment for the case of vacuum $I$ spin vortices only.  The moduli appearing in vacuum $II$ are difficult to analyze due to the coupling of all translational and orientational modes, and give little additional physical information.  We will content ourselves with a brief topological analysis of the vacuum $II$ case, leaving the details for future work.

It is best to consider topological arguments at first to provide a guide in the search for the gapless excitations appearing on the vortex core.  We begin with vacuum $I$ where the non-zero value of the spin field $\chi_i$ breaks the translation symmetry in the $z$ direction.  The global symmetry group of the lagrangian (\ref{Lagrangian}) is given by

\begin{equation}
G_{\rm global} = SO(3)_J \times T
\end{equation}
where $T$ represents the three dimensional translational symmetry.  In vacuum $I$ the ground state is invariant to the symmetry group
\begin{equation}
H_I = U(1)_{J'_z} \times T_{xy} \subset G_{\rm global}.
\end{equation}
However, in the asymptotic limit $r \rightarrow \infty$ the vortex posseses complete axial symmetry
\begin{equation}
H'_I = U(1)_{J_z} \times T_{xy}.
\end{equation}
Gapless excitations of the vortex must appear from generators of $H'_I$ that are broken in the core.  All other modes appearing from the generators of $G_{\rm global}$ would have infinite tension in the kinetic terms.  They would not be normalizable.  

In the previous section we found vortex solutions by assuming rotational symmetry about the $z$-axis.  Thus, since the vortex solutions automatically break the translations in the $x$ and $y$ directions, our vortex solutions obey the symmetry group
\begin{equation}
H'_{I, {\rm vortex}} = U(1)_{J_z}.
\end{equation}
Thus, the degernacy space of the vortex solutions in vacuum $I$ follow from
\begin{equation}
H'_I/H'_{I,{\rm vortex}} =( U(1)_{J_z} \times T_{xy})/U(1)_{J_z} = T_{xy}.
\end{equation}
We should thus find two moduli fields $\xi_{x,y}(t,z)$ associated with the broken translational symmetries by the vortex axis.  These moduli will however acquire a mass gap $m \sim 1/R$ due to the IR cutoff imposed at large distances $R$ from the vortex core.  The details will be shown below.

For the case of vacuum $II$ the cholesteric structure does not appear in the vacuum and thus the full symmetry group of vacuum $II$ is given by:
\begin{equation}
H_{II} = G_{\rm global} = SO(3)_J \times T.
\end{equation}
As in the case of vacuum $I$ the vortex preserves the subgroup $SO(2)_{J_z}$ thus the degeneracy space is:
\begin{equation}
H_{II}/H_{II,{\rm vortex}} = (SO(3)_J \times T)/U(1)_{J_z} = S_{J_\perp}^2 \times T.
\end{equation}
In this case the cholesteric structure only appears in the vortex core and thus an additional longitudinal translational mode appears on the vortex.  The expansion of the low energy Lagrangian in terms of the gapless modes introduces additional interactions between the translational and orientational modes.  We have omitted this topic for sake of clarity and space.

We proceed with the analysis by expanding the Lagrangian by the small adiabatic translations $\xi_{x,y}(t,z)$ of the vortex solution
\begin{align}
&\phi(\vec{x}) \rightarrow \phi(\vec{x}-\vec{\xi}(t,z)) \notag \\
&A_\mu(\vec{x}) \rightarrow A_\mu(\vec{x}-\vec{\xi}(t,z)) \notag \\
&\vec{\chi}(\vec{x}) \rightarrow \vec{\chi}(\vec{x}-\vec{\xi}(t,z)).
\end{align}
Here $\vec{\xi}(t,z)$ only oscillates along the directions $x$ and $y$ perpendicular to the vortex axis.  Proceding with the expansion of (\ref{EnergyDensityEta}) we arrive at the low energy theory
\begin{equation}
\delta^2 E = S_{ab} \partial_z\xi_a \partial_z \xi_b + A\epsilon_{ab} \xi_a \partial_z \xi_b,
\label{LowEnergyVacI}
\end{equation}
where $a$ and $b$ are indices for the $x$ and $y$ directions.  The symbol $\epsilon_{ab}=-\epsilon_{ba}$ is the two dimensional antisymmetric symbol. The parameters $S_{ab}$ and $A$ are given by
\begin{align}\label{dom1}
&S_{ab} = \int d^2x_{\perp} \partial_a \chi_i \partial_b \chi_i, \notag \\
&A = \frac{1}{2}\epsilon_{ab}\int d^2x_\perp \left(2 \partial_z\partial_a \chi_i \partial_b \chi_i+\eta \epsilon_{nm} \partial_a \chi_n \partial_b \chi_m \right).
\end{align}
Here $m$ and $n$ run over the $x$ and $y$ indices. We point out that $S$ and $A$ are in general functions of the $z$ coordinate, owing to the dependence of the spin field $\chi^i$ on $z$.

The expansion in (\ref{LowEnergyVacI}) considered only the kinetic terms in $\xi_a(t,z)$.  We hasten to point out that this expansion is only valid up to the infrared regularization of our vortex solution far from the core.  However, due to the global nature of the spin vortex, the hard cutoff of the vortex solution in the infrared leads to a non-local boundary contribution to the low energy theory.  The infrared cutoff explicitly breaks the translational symmetry in the $x$ and $y$ directions and thus our low energy theory should include mass terms proportional to the inverse cutoff $m \sim 1/R$:
\begin{equation}
\delta^2 \mathcal{L}_{\rm IR} = S_{ab}\dot{\xi}_a\dot{\xi}_b-S_{ab} \partial_z\xi_a \partial_z \xi_b - A\epsilon_{ab} \xi_a \partial_z \xi_b-m^2 \xi_a \xi_a.
\label{LowEnergyVacIIR}
\end{equation}
The details of the calculation of $m^2$ are given in the appendix.  For a cylindrical boundary we find that 
\begin{equation}
m^2 = -\pi\frac{\chi_0^2}{R^2},
\end{equation}
and hence the mass term is negative implying an instability of the vortex solution.  We should point out however, that this value for $m^2$ is dependent on the shape of the boundary.  For non-rotationally symmetric boundaries, the mass term will in fact be tensorial:
\begin{equation}
m^2 \rightarrow m^2_{ab}.
\end{equation}
We consider only the rotationally symmetric boundary.

For relativistic lagrangians this instability means the vortex wants to escape the finite volume.  For non-relativistic cases, the negative mass implies a threshold for the dispersion relations where the Kelvin modes change their direction of propagation \cite{Kobayashi:2013gba}.

For the equation (\ref{LowEnergyVacIIR}) the more appropriate context to consider is the relativistic case - although this is not strictly a relativistic system because the third term in (\ref{LowEnergyVacIIR}) is not Lorentz invariant. There are two important limits in this system characterised by the wavelength of the modes $\lambda$ compared to that of the cholesteric structure $k^{-1}$. In the limit in which $\lambda \gg 1/ k$ the asymptotic values of the $\chi^i$ field dominates the integral appearing in (\ref{dom1}),  the $z$ dependence of $S_{ab}$ disappears and we have that $A=0$ identically. The long wavelength modes effectively ignore the ``high" $k$ structure of the vortex.  As $\lambda \sim 1/k$ the wavelengths become comparable and we may no longer ignore the $z$ dependence of $S_{ab}$. Clearly, $A\neq0$ in this limit. This region becomes difficult to tackle analytically and we won't consider it further. We may however approximate an intermediate region in which $\lambda >1/ k$ where the $z$ dependence of $S_{ab}$ and $A$ is small but $A\neq0$. Then,  varying (\ref{LowEnergyVacIIR}) and using the isotropicity of the matrix $S$ ($S_{xx}=S_{yy}$ and $S_{xy} = S_{yx}$)  we obtain the equation
\be\label{modes}
S_{ac}\left(\xi_a''-\ddot{\xi}_a\right)+A\epsilon_{ac}\xi'_a - m^2\delta_{ac} \xi_a =0.
\ee

Taking the ansatz
\be
\xi_a = C_1 e^{i(\tilde{k}z+\omega t)}  \tilde{\xi}_a,
\ee
where $\tilde{\xi}_a$ is a constant vector, we may reduce this to 
\be
M_{ab}\tilde{\xi}_b =0,
\ee
with 
\be
M_{ab}=\left(S_{ab}(\omega^2-\tilde{k}^2) + A i \epsilon_{ab} \tilde{k}-m^2\delta_{ab} \right),
\ee
for the vector components of $\tilde{\xi}_a$. Finding the corresponding modes is therefore equivalent to finding the zero eigenvalues of the matrix $M_{ab}$. The eigenvalues are
\be
\omega^2_{\pm} = (\omega^2-k^2)S_{yy} - m^2 \pm\sqrt{(A \tilde{k})^2+\left((\tilde{k}^2-\omega^2)S_{yx}\right)^2},
\ee
setting these to zero gives four solutions for $\omega$,
\be
\omega_{1}^2 = \frac{\tilde{k}^2\left(S_{yx}^2-S_{yy}^2\right)-m^2 S_{yy} +\sqrt{(A \tilde{k} S_{yy})^2+S_{yx}^2\left(m^4-(A\tilde{k})^2\right)}}{S_{yx}^2-S_{yy}^2}
\ee
\be
\omega_{2}^2 = \frac{\tilde{k}^2\left(S_{yx}^2-S_{yy}^2\right)-m^2 S_{yy} -\sqrt{(A \tilde{k} S_{yy})^2+S_{yx}^2\left(m^4-(A\tilde{k})^2\right)}}{S_{yx}^2-S_{yy}^2}.
\ee

The (un-normalized) eigenvectors are
\be
\xi_\pm = \left(\pm \frac{i \sqrt{(A\tilde{k})^2+((\tilde{k}^2-\omega^2)S_{yx})^2}}{A\tilde{k}-i(\tilde{k}^2-\omega^2)S_{yx}}\right)
\ee
which define the diagonal basis as
\be
\xi_{\rm diag}=\left(-\frac{A\tilde{k}-(\tilde{k}^2-\omega^2)S_{yx}}{\sqrt{-(A\tilde{k})^2+(\tilde{k}^2-\omega^2)^2S_{yx}^2}}(\tilde{\xi}_x-\tilde{\xi}_y),\;\; \tilde{\xi}_x+\tilde{\xi}_y\right).
\ee

The limit $\lambda \gg 1/k$ (this $k$ should not be confused with $\tilde{k}$) corresponds to
\be
S_{ab} \rightarrow \pi \chi_0^2 \log{\frac{R}{r_0}}\delta_{ab}, \mbox{ and } A \rightarrow 0.
\ee
where $r_0$ represents the coherence length of the vortex core.  In this limit the system is Lorentz invariant and we find the dispersion relation:
\be
\omega = \pm \sqrt{\tilde{k}^2-\Delta^2}, \mbox{ where } \Delta^2 = \frac{1}{R^2\log{R/r_0}}.
\ee
Here $2\pi/\Delta \equiv \lambda_c$ is the critical wavelength of stability.  Modes with wavelengths longer than $\lambda_c$ have purely imaginary $\omega$ and are thus tachyonic \footnote{The ``trivial" solution in which the imaginary part of $\omega$ has a sign corresponding to decaying modes represents a short-lived decaying mode in which the vortex core remains in its original position.}.  For the specific case $\tilde{k} = 0$ the mode represents a complete transverse translation of the vortex solution.  Thus the vortex tends to escape from the finite volume.  On the other hand wavelengths shorter than $\lambda_c$ are real and thus represent stable solutions for the vortex \cite{Kobayashi:2013gba}.  

In the proximicity of the long wavelength limit, where we may still consider equation (\ref{modes}) as a good approximation for the low energy modes, higher order terms in the $S_{ab}$ and $A$ calculation need to be included. Even though we don't perform this computation explicitly we note the main important feature of having a non-zero $A$: it introduces terms in the dispersion relation of the form $\omega\propto \sqrt{\tilde{k}}$, which is characteristic of surface waves in deep water. Finally, leaving the long-wavelength limit entirely equation (\ref{modes}) can no longer be trusted and one is forced to resort to numerics. We leave this treatment to a future publication.

\section{Conclusions}

The goal of this paper was to consider an extension of the analysis presented in \cite{Peterson:2014nma} where mass vortices with a $U(1)_{\rm gauge}$ charge were considered in systems with cholesteric vacuum structure.  Such vortices play a role in many contexts including supersymmetric solitons \cite{Shifman:2009}, type II superconductors \cite{Carlson:2003}, and superfluid $^3$He \cite{Salomaa:1987}.  In the current paper, we extended the analysis to include a different type of vortex, which supported a global topological $U(1)_{J''_z}$ charge in addition to the $U(1)_{\rm gauge}$ charge.  Spin vortices of a similar description were first observed in superfluid $^3$He \cite{Kondo:1992}.  However, for the cholesteric vacuum state we have discussed above, the equivalence of rotations and translations along the $z$ direction in the vacuum leads to a reduction of the degeneracy space $SO(3)_J \rightarrow U(1)_{J''_z}$.

Due to the cholesteric behavior of the vacuum $I$ solution, the $U(1)_{J''_z}$ vortices are stable to deformation \cite{Radzihovsky:2011}.  The low energy effective field theory of these vortices thus gives rise to a $1+1$ dimensional field theory of translational excitations.  A similar result is true in vacuum $II$, however the solutions here are free to tunnel to solutions with no $U(1)_{J''_z}$ charge since the asymptotic form of the vacuum has no cholesteric structure.  We have left a calculation of the tunneling process to future work.

We have focused our attention on the classical description of the $1+1$ dimensional effective field theory of gapless excitations. Specifically, we've calculated the low energy Lagrangan for Kelvin modes of vortices in vacuum $I$ in the long-wavelength limit.  For a relativistic dispersion relation a classical analysis is all that is necessary to count Goldstone modes appearing on the vortex.  However, for the non-relativistic despersion relation, the quantization of modes becomes more complicated.  Some general results for non-Lorentz invariant systems are discussed in \cite{Watanabe:2012}, and to systems with broken spacetime symmetries in \cite{Low:2001}, however we have omitted a detailed analysis here.  We have however explored the effects of a finite volume on the gapless excitations appearing on the vortex as was first pointed out in \cite{Kobayashi:2013gba}.  In particular, the global behavior of the vortex solution in vacuum $I$ lifts the translational degeneracy, and mass gaps are generated for the Kelvin excitations.  

For vacuum $II$ additional orientational modes appear on the vortex, however the details of these modes have added complications and we have left their analysis to future projects.

\section*{Acknowledgments}
The work of A.P. is supported by the Doctoral Dissertation Fellowship at the University of Minnesota. The work of M.S. is supported in part by DOE Grant Number De-Sc0011842. G.T. is funded by Fondecyt Grant Number 3140122. The Centro de Estudios Cient´ıficos (CECS) is funded by the Chilean Government through the Centers of Excellence Base Financing Program of Conicyt.

\subsection*{Appendix}
In this appendix we wish to show the details of the calculation of the boundary terms leading to the mass term appearing for the translational moduli in (\ref{LowEnergyVacIIR}).  We will first show that the linear term in $\xi_i$ vanishes, confirming that no instabilities appear on the vortex.  Following this calculation we will expand to second order in $\xi_i$ and determine the mass term for the particular vortex configuration.

We begin with a calculation of the linear term for the expansion parameter $\xi$, which we may take to be constant.  The $t$ and $z$ dependence is accounted for in the kinetic terms shown in (\ref{LowEnergyVacI}).  To first order in $\delta \chi_i = -\xi_a \partial_a \chi_i$
\begin{eqnarray}
\delta E &=& 2\partial_i (\delta \chi_j \partial_i \chi_j)+\eta \varepsilon_{ijk}(\delta\chi_i \partial_j \chi_k+ \chi_i \partial_j \delta\chi_k)+\delta \chi_i \frac{\partial V}{\partial \chi_i} \notag \\
&=& 2\partial_i \left(\delta \chi_j \partial_i \chi_j -\frac{\eta}{2} \varepsilon_{ijk}\chi_j \delta\chi_k\right) \notag \\
&&+2\delta\chi_i \left(-\partial^2 \chi_i +\eta \varepsilon_{ijk}\partial_j\chi_k +\frac{1}{2}\frac{\partial V}{\partial \chi_i}\right) \notag \\
&=&-2\xi_a\partial_i\left(\partial_a\chi_j \partial_i \chi_j + \frac{\eta}{2}\varepsilon_{ijk}\partial_a\chi_j \chi_k \right),
\end{eqnarray}
where the second line in the second equality vanished by the equations of motion for $\chi_i$.  To obtain the contribution of the linear term to the low energy theory, we integrate over the perpendicular directions $x$ and $y$:
\begin{eqnarray}
\delta E &\rightarrow& -2\xi_a \int d^2x_\perp \partial_i\left(\partial_a\chi_j \partial_i \chi_j + \frac{\eta}{2}\varepsilon_{ijk}\partial_a\chi_j \chi_k \right) \notag \\
&=&-2\xi_a \int d\theta \; R \; \hat{r}_i\left(\partial_a\chi_j \partial_i \chi_j + \frac{\eta}{2}\varepsilon_{ijk}\partial_a\chi_j \chi_k \right),
\end{eqnarray}
where we have used Greens theorem, and $\hat{r}$ is the unit vector in the radial direction.

At this point we may substitute the asymptotic form of $\chi_ i$.  Ignoring the $z$ dependence 
\begin{equation}
\chi_i \rightarrow \chi_0 \frac{x_i}{r} \mbox{ and } \partial_j \chi_i \rightarrow \left(\frac{\delta_{ij}}{r}-\frac{x_i x_j}{r^3}\right) 
\label{AsymptoticForm}
\end{equation} 
Inserting this into the equation above we find that the linear term identically vanishes as required.

We now proceed to calculate the second order contribution to the energy density
\begin{eqnarray}
\delta^2E &=& \partial_j \delta\chi_i \partial_j \delta\chi_i+\eta \varepsilon_{ijk}\delta\chi_i \partial_j \delta\chi_k +\frac{1}{2}\frac{\partial^2V}{\partial\chi_i\partial\chi_j}\delta\chi_i\delta\chi_j \notag \\
&=& \partial_j\left(\delta\chi_i \partial_j\delta\chi_i\right) +\delta\chi_i\left(-\partial^2\delta\chi_i+\eta \varepsilon_{ijk}\partial_j\delta\chi_k+\frac{1}{2}\frac{\partial^2V}{\partial\chi_i\partial\chi_j}\delta\chi_j\right). \notag \\
\end{eqnarray}
Inserting $\delta\chi_i = -\xi_a \partial_a \chi_i$
\begin{eqnarray}
\delta^2E &=& \xi_a \xi_b \partial_i\left(\partial_a \chi_j \partial_i\partial_b \chi_j\right) \notag \\
&&+ \xi_a\xi_b\partial_a\chi_i\partial_b\left(-\partial^2\chi_i +\eta\varepsilon_{ijk}\partial_j\chi_k+\frac{1}{2}\frac{\partial V}{\partial\chi_i}\right) \notag \\
&=&\xi_a\xi_b\partial_i\left(\partial_a\chi_j \partial_i\partial_b\chi_j\right),
\end{eqnarray}
where we have used the equation of motion in the last line.

We may now integrate over the perpendicular coordinates and use Green's theorem:
\begin{equation}
\delta^2E \rightarrow \xi_a \xi_b \int d^2x_\perp \partial_i\left(\partial_a \chi_j \partial_i \partial_b \chi_j\right)=\xi_a \xi_b \int d\theta R\;\hat{r}_i \partial_a\chi_j \partial_i \partial_b\chi_j
\end{equation}
Finally, we insert the asymptotic form of $\chi_i$ in (\ref{AsymptoticForm}) and find
\begin{equation}
\delta^2E = -\frac{\pi\chi_0^2}{R^2}\xi_a\xi_a.
\end{equation}
This particular value of $m^2$ was determined from our particular choice of boundary contour.  Depending on the boundary conditions, this term may be modified and may even be tensorial depending on the symmetries of the contour choice.  However, by dimensional analysis we are guaranteed to have
\begin{equation}
m^2_{ab} \sim -\frac{\chi_0^2}{R^2}.
\end{equation}

\end{document}